\documentclass[12pt]{article}
\usepackage[T1]{fontenc}
\usepackage{amsfonts,amssymb,amsmath,bbm}
\usepackage[normalem]{ulem}
\usepackage{graphicx}
\usepackage{color}
\usepackage{amsmath,amsfonts}
\graphicspath{ {./graphs/} }
\usepackage{subcaption}
\usepackage{multirow}
\usepackage{url}
\usepackage{hyperref}
\usepackage{tabularx}
\usepackage{algorithm}
\usepackage{algpseudocode}
\hypersetup{colorlinks,linktocpage=true}
\newtheorem{prop}{Proposition}

\hypersetup{colorlinks,%
citecolor=blue,%
filecolor=blue,%
linkcolor=red,%
urlcolor=blue,%
}

\DeclareMathOperator*{\argmin}{arg\,min}

\begin{document}

\date{}

\title{Cycles Communities from the Perspective of Dendrograms and Gradient Sampling}
\author{
Sixtus Dakurah\\
University of Wisconsin-Madison\\
\textit{sdakurah@wisc.edu}
}

\maketitle

\begin{abstract}
Identifying and comparing topological features, particularly cycles, across different topological objects remains a fundamental challenge in persistent homology and topological data analysis. This work introduces a novel framework for constructing cycle communities through two complementary approaches. First, a dendrogram-based methodology leverages merge-tree algorithms to construct hierarchical representations of homology classes from persistence intervals. The Wasserstein distance on merge trees is introduced as a metric for comparing dendrograms, establishing connections to hierarchical clustering frameworks. Through simulation studies, the discriminative power of dendrogram representations for identifying cycle communities is demonstrated. Second, an extension of Stratified Gradient Sampling simultaneously learns multiple filter functions that yield cycle barycenter functions capable of faithfully reconstructing distinct sets of cycles. The set of cycles each filter function can reconstruct constitutes cycle communities that are non-overlapping and partition the space of all cycles. Together, these approaches transform the problem of cycle matching into both a hierarchical clustering and topological optimization framework, providing principled methods to identify similar topological structures both within and across groups of topological objects.
\end{abstract}

\section{Introduction}
In topological data analysis (TDA), understanding the structural similarities and differences between topological objects remains a fundamental challenge with applications spanning neuroscience, network analysis, and shape comparison \cite{dakurah2025discrete,dakurah2022modelling,dakurahregistration,garcia2022fast,leygonie2023gradient,reani2021cycle}. In this work, the focus is on the {\em cycle} structure of topological objects and the signals defined on them. Cycles, or 1-dimensional holes in simplicial complexes, capture fundamental topological invariants that are robust to noise and deformation \cite{dakurah2024subsequence,dakurah2025topologically}. The central question addressed here is how to identify and group similar cycles within and across topological objects, what we term cycle communities. This notion is motivated by the observation that not all cycles in a topological object are equally similar: some naturally group together based on their topological persistence properties, while others remain distinct. Understanding this community structure provides a new lens for analyzing complex topological structures.

Two complementary approaches to cycle community construction are developed. The first approach leverages the hierarchical structure inherent in persistent homology through dendrogram representations. Dendrograms provide a natural framework for visualizing and analyzing the nested relationships among homology classes as they appear and disappear across filtration values \cite{schweinhart2015statistical,obayashi2018volume,gueunet2017task,morozov2013distributed}. By constructing dendrograms from persistence intervals using merge-tree algorithms \cite{obayashi2018volume,schweinhart2015statistical} and comparing them using metric structures such as the Wasserstein distance on merge trees \cite{pont2021wasserstein,pegoraro2021metric}, topologically similar cycles can be identified and differences in cycle structure across networks can be assessed. This dendrogram-based methodology offers an interpretable, tree-structured view of cycle communities that directly reflects the birth-death dynamics of topological features encoded in persistence diagrams \cite{edelsbrunner2008persistent,zomorodian2005computing}. Moreover, the connection between dendrograms and hierarchical clustering \cite{nielsen2016hierarchical,murtagh2012algorithms,day1984efficient,lee2011computing} enables leveraging well-established statistical frameworks for comparing topological objects, including comparison against baseline distance measures such as the cophenetic correlation coefficient \cite{sokal1962comparison}, Gromov-Hausdorff distance, and bottleneck distance \cite{murtagh2012algorithms,lee2011computing}.

The second approach extends the Stratified Gradient Sampling (SGS) procedure \cite{leygonie2023gradient}, which was recently introduced for registering topological objects to a common template by learning a {\em filter function} that can faithfully recover all topological objects registered to it. This framework has proven particularly effective in applications to functional brain networks, where identifying corresponding cycles across subjects is essential for understanding common neural pathways \cite{dakurah2025spanning,dakurah2025brain,dakurah2022modelling,dakurahregistration}. Here, the SGS procedure is extended from learning a single filter function to learning a collection of filter functions simultaneously. This collection yields cycle barycenter functions, each of which can faithfully reconstruct a set of cycles. The set of cycles each filter function can reconstruct constitutes cycle communities, where these communities are non-overlapping and partition the space of all cycles. This extension requires careful theoretical development, including establishing the smooth stratifiability of the objective functions \cite{bolte2007clarke,leygonie2021framework} and developing the optimization procedure through stratified gradient sampling on nonsmooth and nonconvex functions \cite{dakurahregistration,leygonie2023gradient}.

Together, these two approaches provide both a descriptive framework through dendrograms and a constructive optimization-based framework through stratified gradient sampling. The dendrogram approach excels at revealing hierarchical relationships and enabling statistical comparison across groups, while the SGS approach offers a principled way to learn community structures through topological registration.

The remainder of this paper is organized as follows. Section~\ref{sec:prelim} provides necessary background on topological concepts including simplicial complexes, filtrations, and persistent homology. Section~\ref{sec:dendrograms} develops the dendrogram-based methodology for cycle community detection and statistical analysis. Section~\ref{sec:sgs} presents the stratified gradient sampling theory and the cycle barycenter construction. Section~\ref{sec:dc} concludes with a discussion of contributions and future directions.

\section{Topological Preliminaries}\label{sec:prelim}

\subsection{Homology}
This section briefly introduces the basic concepts in algebraic topology that will be pertinent to this work. We start with the definition of a simplicial complex.

\subsubsection{Simplicial complex}

A $k$-simplex $s_i^k = (v_{i_0}, \cdots, v_{i_k})$ is a $k$-dimensional convex hull (polytope) of nodes $v_{i_0}, \cdots, v_{i_k}$.
A simplicial complex $K$ is a set of simplices such that for any $s_i^k, s_j^k \in K$,
  $s_i^k \cap s_j^k$ is a face of both simplices; and a face of any simplex $s_i^k \in K$ is also a simplex in $K$ \cite{edelsbrunner2008persistent}. 
  A 0-skeleton is a simplicial complex consisting of only nodes. A 1-skeleton is a simplicial complex consisting of nodes and edges. Graphs are 1-skeletons.
A $k$-chain is a finite sum $\sum a_i s_i^k$, where the $a_i$ are either 0 or 1. The set of $k$-chains forms a group and a sequence of these groups is called a chain complex. To relate different chain groups, we use the boundary maps \cite{topaz2015topological}. For two successive chain groups $\mathcal{K}_k$ and $\mathcal{K}_{k-1}$, the boundary operator $\partial_k: \mathcal{K}_k \longrightarrow \mathcal{K}_{k-1}$ for each $k$-simplex $s_k$ is given by
$$
    \partial_k(s_i^k) = \sum_{l=0}^k(-1)^l(v_{i_0}, \cdots, \widehat{v_{i_l}}, \cdots, v_{i_k})
    \label{eqn:boundary_map},
$$
where $(v_{i_0}, \cdots, \widehat{v_{i_l}}, \cdots, v_{i_k})$ gives the $k$-$1$ faces of $s_i^k$ obtained by deleting node $\widehat{v_{i_l}}$.
The kernel of the boundary operator is denoted as $\mathcal{Z}_k = ker(\partial_k)$ and its image denoted as $\mathcal{B}_{k} = img(\partial_{k+1})$. $\mathcal{Z}_k$ and $\mathcal{B}_{k}$ are the subspaces of $\mathcal{K}_k$. The elements of $\mathcal{Z}_k$ and $\mathcal{B}_{k}$  are known as $k$-cycles and $k$-boundaries respectively \cite{hatcher2002algebraic}. Note that $\mathcal{B}_{k} \subseteq \mathcal{Z}_k$. The set quotient $\mathcal{H}_k = \mathcal{Z}_k/\mathcal{B}_{k}$ is termed as the $k$-th homology group \cite{chen2010,hatcher2002algebraic,topaz2015topological}. The $k$-th Betti number $\beta_{k} = rank(\mathcal{H}_k)$ counts the number of algebraically independent $k$-cycles. In this work, our domain will be exclusively 1-skeletons (graphs). This implies the first homology group is $\mathcal{H}_1 = ker(\partial_1)$ since $img(\partial_2) = \varnothing$. Hence $\mathcal{H}_1$ is a vector space of $1$-cycles and similarly $\mathcal{H}_0$ is a vector space of $0$-cycles. The $k$-th homology group contains important information about the birth and death of a homology class. The mechanism for tracking the birth and death of these classes is often formulated in the context of filtration \cite{chung2015persistent,edelsbrunner2008persistent,zomorodian2005computing}.

\subsection{Filter Functions and Persistent Homology}
The common approach to studying the persistence of homology generators is to construct a filtration on the set of $0$-simplices (nodes). We adopt a slightly different approach in this work and define the filtration along the $1$-simplices. By adopting this approach, we assume the homology generators are dependent on the connection relation (correlation, distance, among others) between any two points in the topological space. To properly develop the theory, we also restrict our exposition to one-dimensional simplicial complex. This stems from the fact that 1-cycles/loops can be fully identified from the one-dimensional simplicial complex. We start with the full simplex which we assume is a graph, and sequentially threshold. 

\subsubsection{Filtration}
\begin{figure}[ht]
 \centering
 \includegraphics[width=1\linewidth,clip=true]{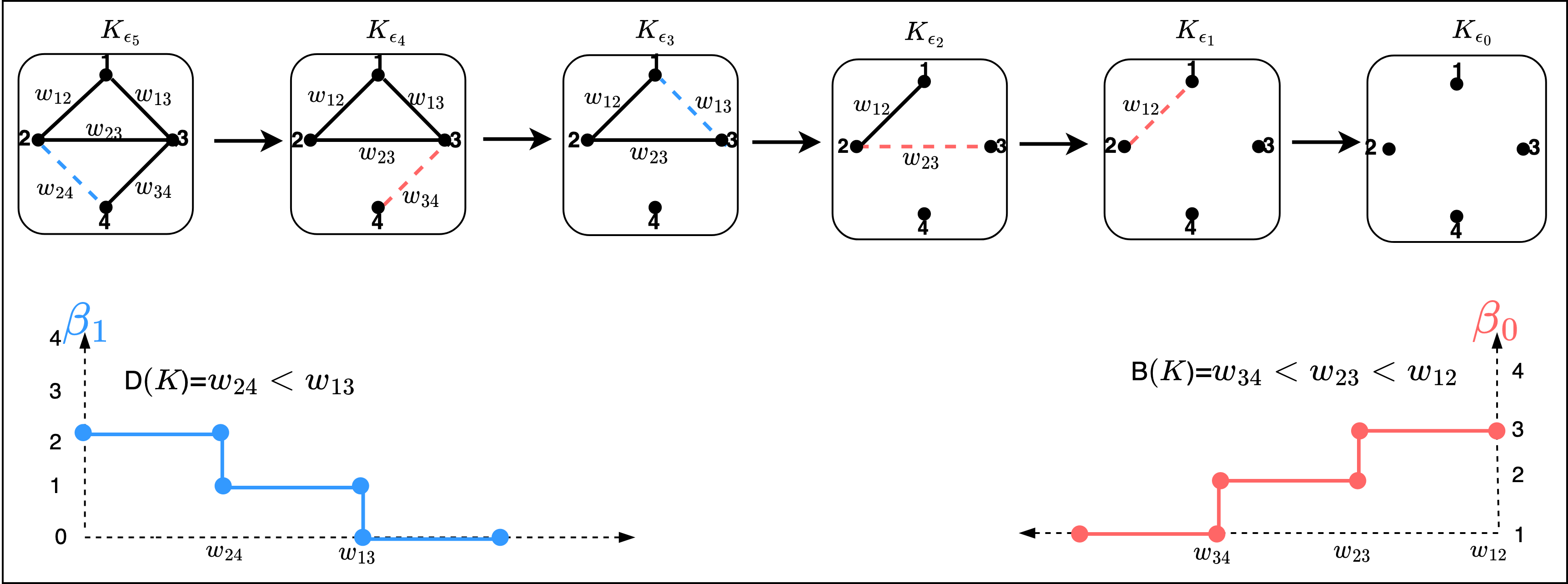}
 \caption{An illustration of the filtration on a 1-dimensional simplicial complex. A dashed red or blue line indicates an edge that has been deleted. From top-left, the full simplicial complex which is sequentially thresholded to the point set(top-right). Bottom-left, the non-increasing count of the number of 1-cycles/loops. Bottom-right, the non-decreasing count of the number of connected components.}
 \label{fig:filtration-report}
\end{figure}

A {\em filtration} defined  on a simplicial complex $K$ is a map $F: K \xrightarrow{} \mathbb{R}^{|\mathcal{K}_1|}$. This map induces an inclusion relation of the simplicial complex $K$ such that $K_{\epsilon_r} \subseteq K_{\epsilon_k}$ whenever $F(\epsilon_k) \ge F(\epsilon_r)$ where $\epsilon_k$ and $\epsilon_r$ are some indexes defined on the $k$-th and $r$-th 1-simplices respectively. $F$ will be termed a { \em filter} function. For example, if the topological space under consideration is a graph, $F(\epsilon_k)$ could be the edge weight associated with that edge (1-simplex). The Figure \ref{fig:filtration-report} illustrates an example filtration over a one-dimensional simplicial complex with four nodes (0-simplices) and five edges (1-simplicies). For example, ${{w}_{24}} < {w}_{34} < w_{13} < w_{23}  < w_{12}$ are ordered thresholds (edge weights). As we sequentially move along these thresholds, more 1-simplices are disconnected, increasing the number of connected components $(\beta_0)$, and decreasing the number of 1-cycles $(\beta_1)$. This increase in $\beta_0$ and decrease in $\beta_1$ is monotonic. To see the inclusion relation with the filter function $F$, observe that $F(\epsilon_5) = w_{24}$, and $F(\epsilon_4) = w_{34}$ which implies that $F(\epsilon_5) > F(\epsilon_4)$. Further observe that $K_{\epsilon_4} \subset K_{\epsilon_5}$ since $K_{\epsilon_4}$ includes all the 1-simplices in $K_{\epsilon_5}$ except that joining point 3 and 4. The filtration allows us to track the birth of connected components (0-cycles) and the death of $1$-cycles over the span of the filtration values. The persistence of a connected component or 1-cycle that appears at filtration value $b_i$ and disappears at filtration value $d_i$ is represented by the interval $\left[b_i,  d_i\right]$. The length of this interval characterizes the persistence (life-span) of the 0-cycles or 1-cycles. This characterization is formalized through the concept of persistent homology and is discussed in the next section. 

\subsubsection{Persistent Homology}
The {\em Persistent Homology} (PH) is a framework for tracking the topological changes of $K$ induced by the {\em filter function} $F$ over the span of the filtration \cite{cohen2009persistent,edelsbrunner2008persistent}. More specifically, it tracks the filtration value $b_i$ at which a cycle appears and the filtration value $d_i$ at which the cycle disappears. 
A finite collection of $\left[b_i,  d_i\right]$ can be summarized in the form of a \textit{barcode} (Figure \ref{fig:filtration-report} bottom). For our filtration framework adopted, when a connected component is born, it never dies, hence it's lifespan is $[b_i, \infty)$.  Similarly, all the 1-cycles are born when the graph is first formed and it's lifespan is represented as $(-\infty, d_i]$. For practical considerations, it is sufficient to replace the infinite values with the minimum and maximum filtration values, and it is the approach that will be adopted in this work. During the filtration, when an edge is deleted, a 0-cycle is formed or a 1-cycle dies. Both events can not occur at the same time. 
The persistence of a cycle that appears at filtration value $b_i$ and disappears at filtration value $d_i$ is given by the interval $\left[b_i,  d_i\right]$. A finite collection of $\left[b_i,  d_i\right]$ can be summarized in the form of a \textit{barcode}. 
Ignoring $\pm \infty$, the collection of birth values $B(\mathcal{X})$ and  death values $D(\mathcal{X})$ can be represented as the $0$D and $1$D \textit{barcode}:
\begin{equation}
    B(K) = b_1  < b_2 < \dots < b_{p},\quad  D(K) = d_1 < d_2 < \dots < d_{q}.
    \label{eqn:barcodes}
\end{equation}
The point $(b_i,d_i) \in \mathbb{R}^2$ is referred to as the {\em persistence interval}, and its length measures the persistence of a given cycle. Longer persistence indicates topological signal whiles shorter persistence might represent topological noise.

\section{Dendrograms of Homology Groups}\label{sec:dendrograms}
The filtration on the simplicial complex in the previous section creates homology classes. We now develop the theory for obtaining a dendrogram representation of these homology classes. We will assume throughout this section that the simplicial complex is finite.

\subsection{Filtration Graphs of the Homology Groups}

The persistence intervals $\{(b_i, d_i)\}$ associated with the homology classes in the groups $\mathcal{H}_k$ can be used to characterize the notion of a  tree structure of the homology classes in $\mathcal{H}_k$ through the concept of merge trees \cite{obayashi2018volume,gueunet2017task,morozov2013distributed}. More specifically, if we let $\mathbb{P}_k(K) = \{(b_i, d_i)\}$, regarded in the literature as a persistent diagram. Then for any two birth-death pairs $(b_i, d_i)$, $(b_j, d_j) \in \mathbb{P}_k(K)$, we can compute the two different $k$-cycles associated with this pair. It has been shown that if the resulting $k$-cycles are optimal with regards to the length of the cycle, 
then $\mathbb{P}_k(K)$ can be regarded as set of trees, referred to as persistence trees \cite{obayashi2018volume,schweinhart2015statistical}. This tree structure is due to the nested nature of the birth-death pair in $\mathbb{P}_k(K)$ and minimality of the resulting cycles is not required. An algorithm for computing this tree structure using the concept of merge trees was first presented in \cite{schweinhart2015statistical}. However, their algorithm is not applicable to homology classes in $\mathcal{H}_1$. This limitation was addressed and the algorithm enhanced to accommodate homology classes of any dimension \cite{obayashi2018volume}. This also has the theoretical limitation that it can not handle homology classes  where the death time is infinite in the generating birth-death pair. Further, the algorithm presumes the cycles in $\mathcal{H}_k$ are optimal in terms of their length. In what follows, we describe an algorithm for constructing the tree associated with classes in $\mathcal{H}_0$ and $\mathcal{H}_1$. This is motivated by the concepts and algorithms in \cite{obayashi2018volume,schweinhart2015statistical,gueunet2017task,morozov2013distributed}, and more specifically an alternative construction presented in \cite{schweinhart2015statistical} if $\mathbb{P}_k(K)$ is already precomputed. To facilitate the construction of the algorithm, consider the following graph structure. Let ${T} = ( V_T, E_T )$ be a graph where $V_T$ is the vertex set and $E_T$ is the set of links connecting any two vertices. Note that we've purposely reserved the terms vertex and links (instead of the nodes and edges used in previous sections) for this particular graph structure. We denote elements of $V_T = \{ s_i^0 \}$ and elements of $E_T$ with $\{s_i^1\}$. Then the algorithm reads:

\begin{algorithm}
\caption{Constructing the trees by merge-tree algorithm}\label{alg:merge-tree}
\begin{algorithmic}
\Require $\mathbb{P}_k(K)$\\
Initialize $V_T = \emptyset$ and $E_T = \emptyset$
\Ensure $0 \le k \le 1$
\If{k = 0}
\State $\mathbf{\epsilon} \gets sort(\{b_1, \cdots, b_p\})$ \Comment{ascending}
%\State $V_T = \{s_i^0: 1 \le i \le p \}$
\ElsIf{k = 1}
\State $\mathbf{\epsilon} \gets sort(\{d_{p+1}, \cdots, d_{p+q}\})$ \Comment{descending}
%\State $V_T = \{s_i^0: 1 \le i \le q \}$
\EndIf
\For{$\mathbf{\epsilon}_i \in \mathbf{\epsilon}$}
\State  $s_i^1 \thicksim \mathbf{\epsilon}_i$ \Comment{associate a link with $\mathbf{\epsilon}_i$} 
\If{$\mathbf{\epsilon}_i  = max(\mathbf{\epsilon})$}
\State $(s_{i_1}^0, s_{i_2}^0) \thicksim s_i^1$ \Comment{associate vertices with $s_i^1$}
\State $V_T = V_T \cup \{s_{i_1}^0, s_{i_2}^0\}$
\Else
\State $(s_{(i-1)_2}^0, s_{i_2}^0) \thicksim s_i^1$ \Comment{associate vertices with $s_i^1$}
\State $V_T = V_T \cup \{s_{(i-1)_2}^0, s_{i_2}^0\}$  
\EndIf
\State $E_T = E_T \cup s_i^1$
\EndFor\\
\noindent
\textbf{Return:} ${T} = (V_T, E_T )$
\end{algorithmic}
\end{algorithm}

The returned graph $T$ is weighted with the link weights been either $b_i$ or $d_i$ of $\mathbb{P}_k(K)$. More specifically, for elements in $\mathcal{H}_0$, the vertices are the $0$-cycles and the links and the link weights are the birth values. Similarly, for elements in $\mathcal{H}_1$, the vertices are the $1$-cycles and the link weights are the death values. This persistence tree $T$ is often transformed into a more informative representation in the form of a dendrogram.

\subsection{Dendrogram Representation of Persistence Trees}

A dendrogram is a diagram representing a tree \cite{nielsen2016hierarchical}. The persistence tree $T$ obtained from Algorithm \ref{alg:merge-tree} can be transformed into a dendrogram. The dendrogram representation highlights the cluster structure in the tree. In fact the dendrogram representation for the persistence trees exactly mimics hierarchical clustering \cite{day1984efficient,murtagh2012algorithms,nielsen2016hierarchical}.
Hierarchical clustering involves building a binary merge tree. Essentially, we maintain an "active set" of clusters and at each stage, decides which two clusters to merge. When two clusters are merged, they are each removed from the active set and their union added to the active set. This iterates until the active set comprises of only a single cluster. This merger process generates a tree which is often referred to as the dendrogram \cite{day1984efficient,murtagh2012algorithms,nielsen2016hierarchical}. In choosing which clusters to merge, the SLC employs the single-linkage criterion, i.e., it merges groups based on the shortest distance over all possible pairs \cite{day1984efficient,murtagh2012algorithms,lee2011computing}. From the tree in Algorithm \ref{alg:merge-tree}, the only information needed to decorate the dendrogram are the cluster labels. Note that the clusters are the vertices in $T$.
We first illustrate the dendrogram construction for $I_0(K)$ for elements in $\mathcal{H}_0$, i.e., $0$-cycles (connected components) using the set of birth values. The dendrogram representation of classes in $\mathcal{H}_0$ exhibits maximal branching whiles that of $\mathcal{H}_1$ exhibits minimal branching \cite{schweinhart2015statistical}. The dendrogram representation is not unique \cite{nielsen2016hierarchical}, and this will be explored in later sections.

\subsubsection{Dendrogram Representation of Connected Components}
\begin{figure}[ht!]
	\centering
	\includegraphics[width =\textwidth]{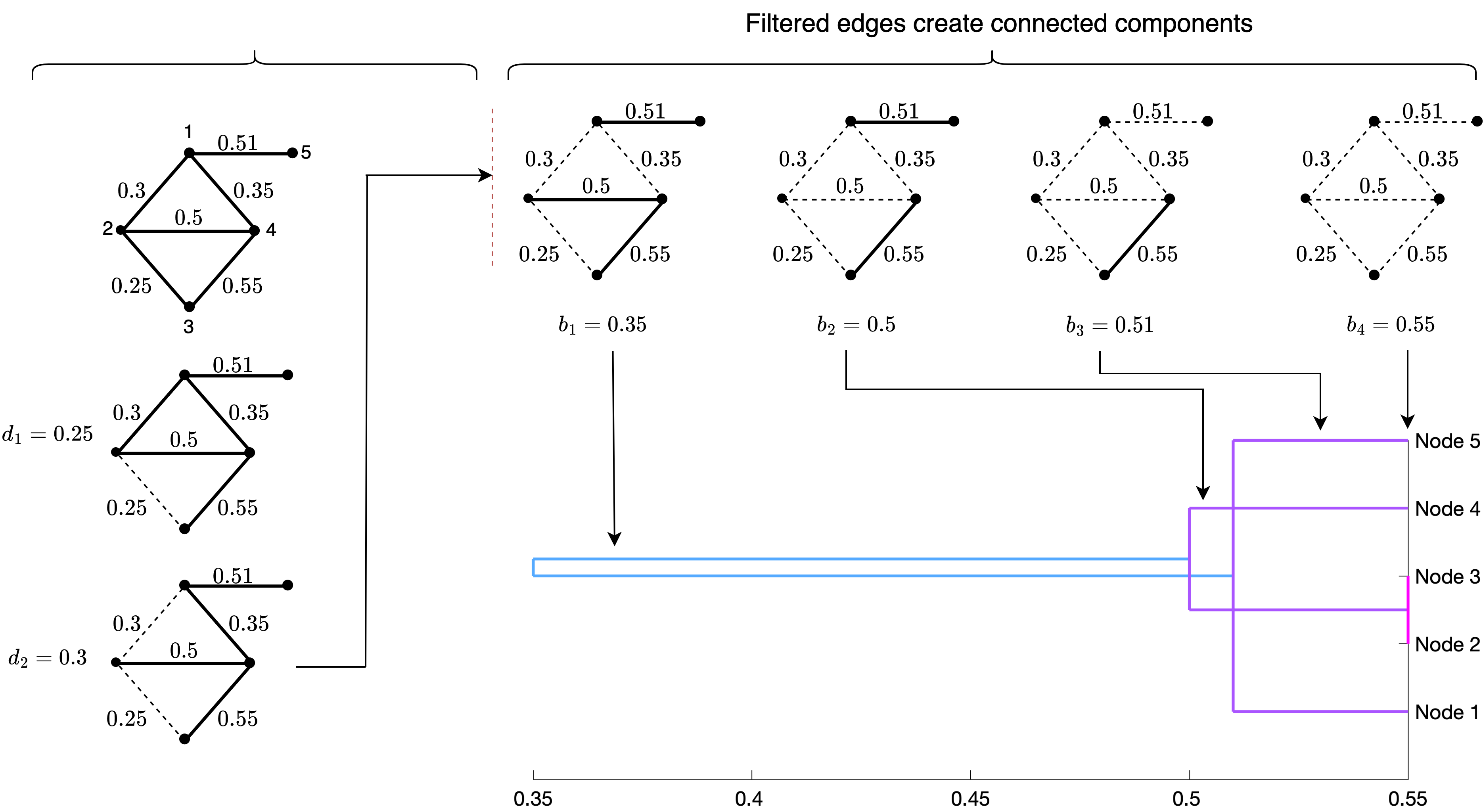}
	\caption{Illustration of the correspondence between the filtration and the dendrogram. Left-top, the five node network. The first two filtration values $d_1 = 0.25, d_2 = 0.3$ destroys the two independent cycles and are not part of the birth set. At the third filtration value denoted $b_1 = 0.35$, two connected components are created when edges with weight at or less than $b_1$ are deleted. The two blue lines in the dendrogram, identifies the two connected components (clusters). The process continues sequentially until we get to the highest filtration value $b_4 = 0.55$, resulting in five connected components (clusters), the node set of the network.}
	\label{fig:dendogram-birth-death}
\end{figure}

Consider the five-node network in the top-left of Figure \ref{fig:dendogram-birth-death}. The first two filtration values $d_1 = 0.25, d_2 = 0.3$ destroys the two independent cycles and are not part of the birth set. At the third filtration value denoted $b_1 = 0.35$, two connected components are created when edges with weight at or less than $b_1$ are deleted. The two blue lines in the dendrogram, bottom-right of Figure \ref{fig:dendogram-birth-death} identifies the two connected components (clusters). The process continues sequentially until we get to the highest filtration value $b_4 = 0.55$, resulting in five connected components (clusters), the node set of the network. Hence at each birth value $b_i$, the dendrogram can uniquely identify elements in the clusters making up the connected components. The above description translates Algorithm \ref{alg:merge-tree} into a dendrogram.

\subsubsection{Dendrogram Representation of 1-Cycles}
\begin{figure}[ht!]
\centering
 \begin{subfigure}[b]{0.32\textwidth}
     \centering
     \includegraphics[width=\textwidth]{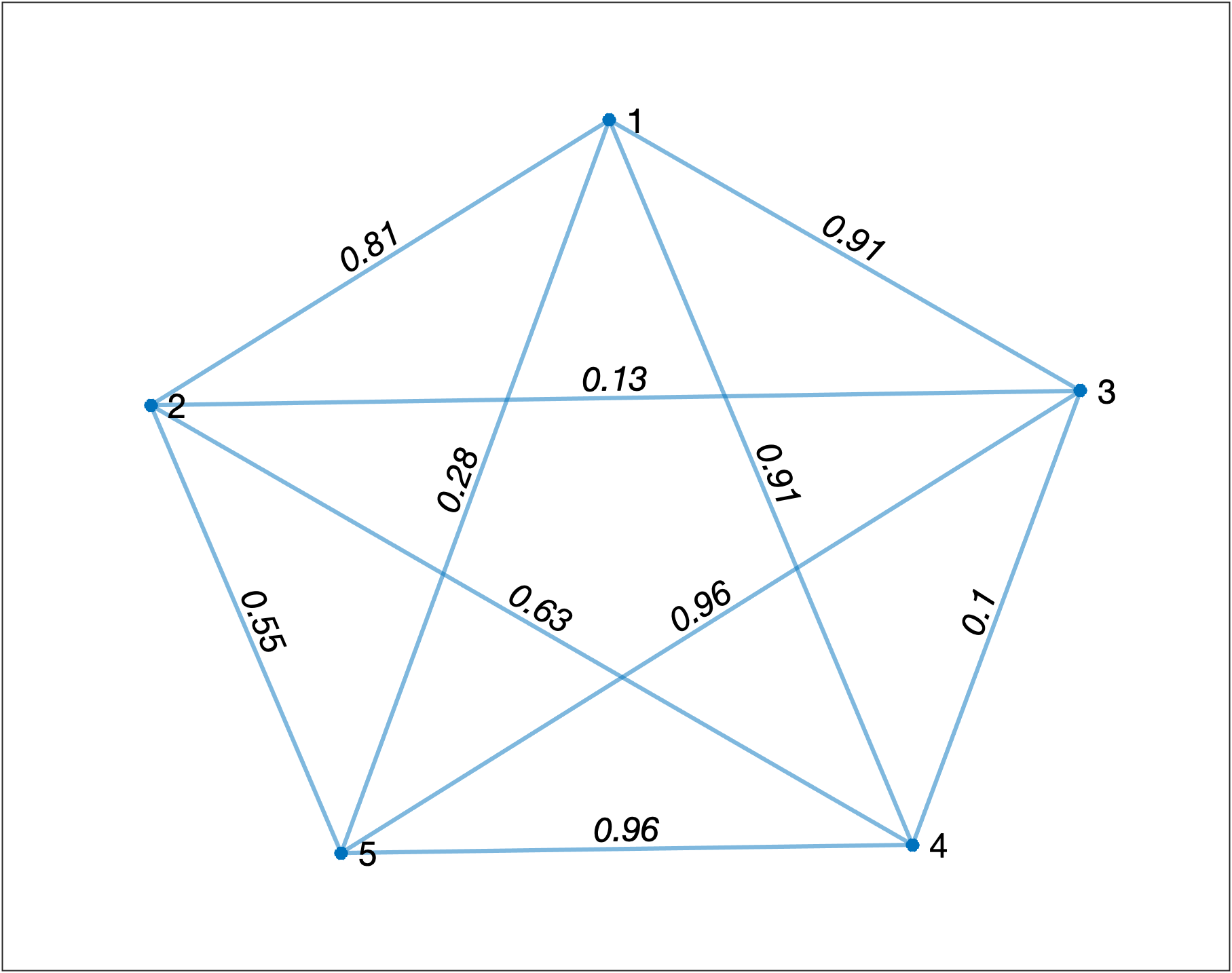}
      \caption{}
     \label{fig:full}
 \end{subfigure}
 \hfill
 \begin{subfigure}[b]{0.32\textwidth}
     \centering
     \includegraphics[width=\textwidth]{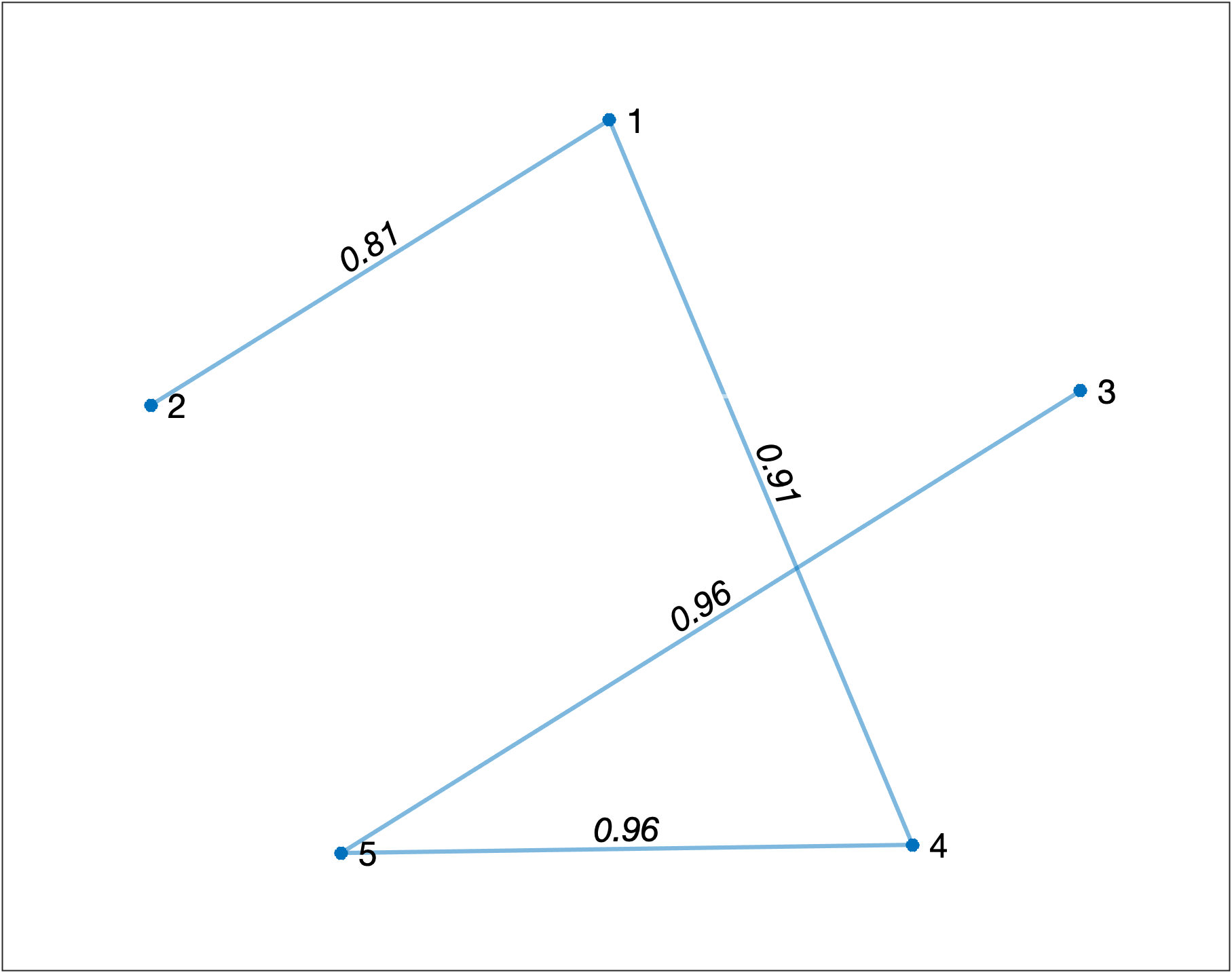}
      \caption{}
     \label{fig:birth}
 \end{subfigure}
  \hfill
 \begin{subfigure}[b]{0.32\textwidth}
     \centering
     \includegraphics[width=\textwidth]{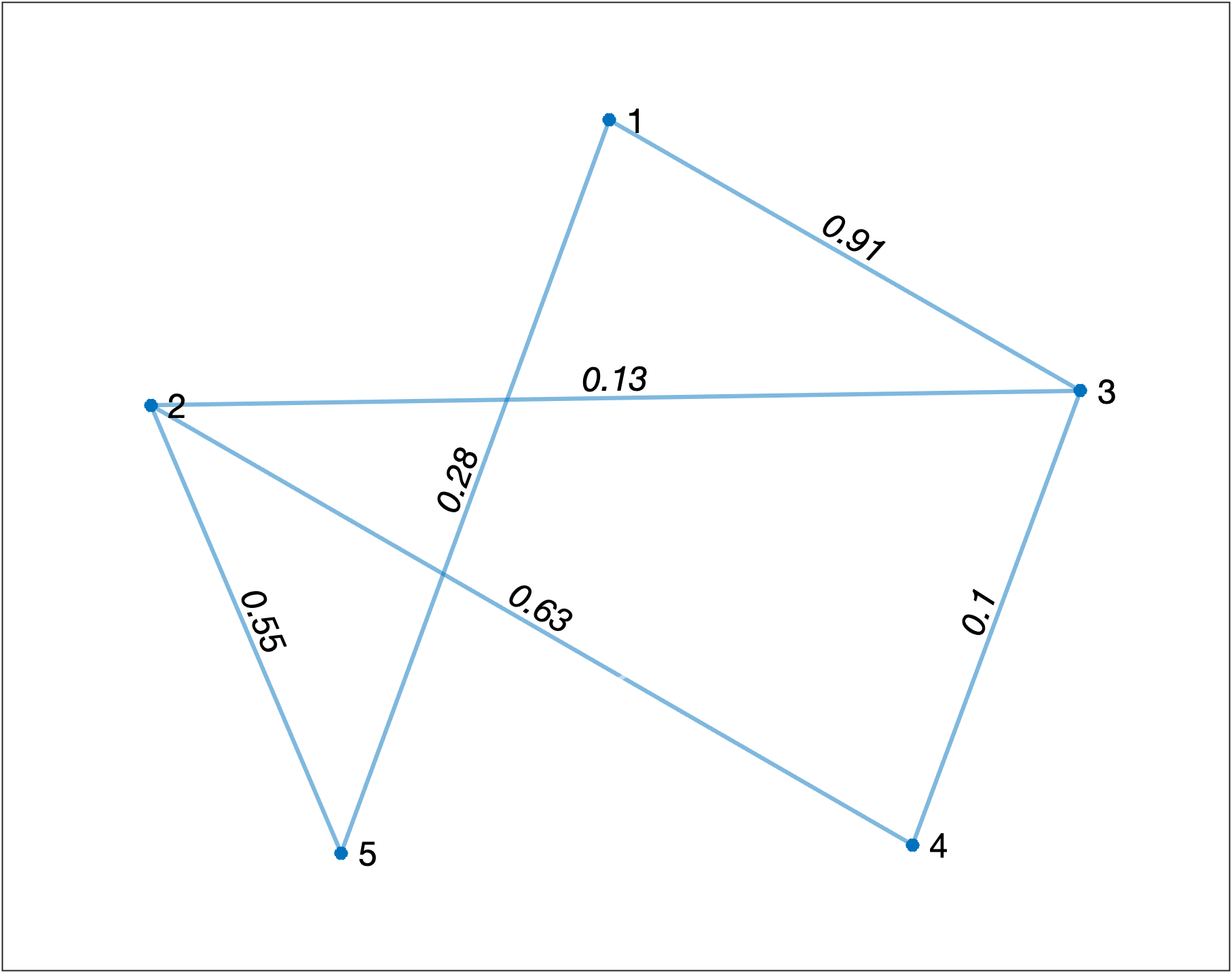}
      \caption{}
     \label{fig:death}
 \end{subfigure}
 \caption{The network and its decomposition. (a) The fully-connected network. (b) The maximum spanning tree. (c) The non-maximum spanning tree.}
\end{figure}

\begin{figure}[ht!]
\centering
 \begin{subfigure}[b]{0.44\textwidth}
     \centering
     \includegraphics[width=\textwidth]{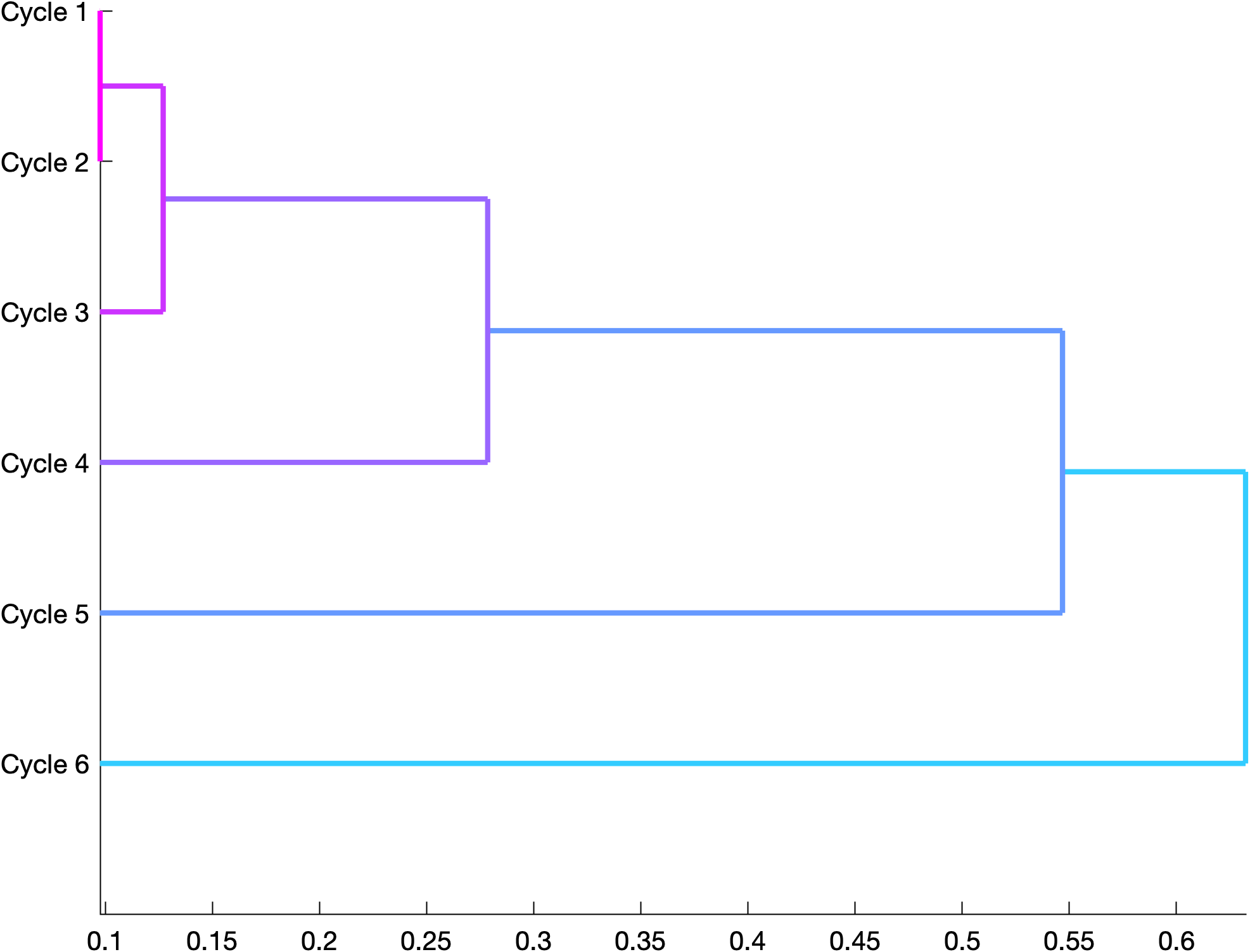}
      \caption{}
     \label{fig:dddeath}
 \end{subfigure}
 \hfill
 \begin{subfigure}[b]{0.44\textwidth}
     \centering
     \includegraphics[width=\textwidth]{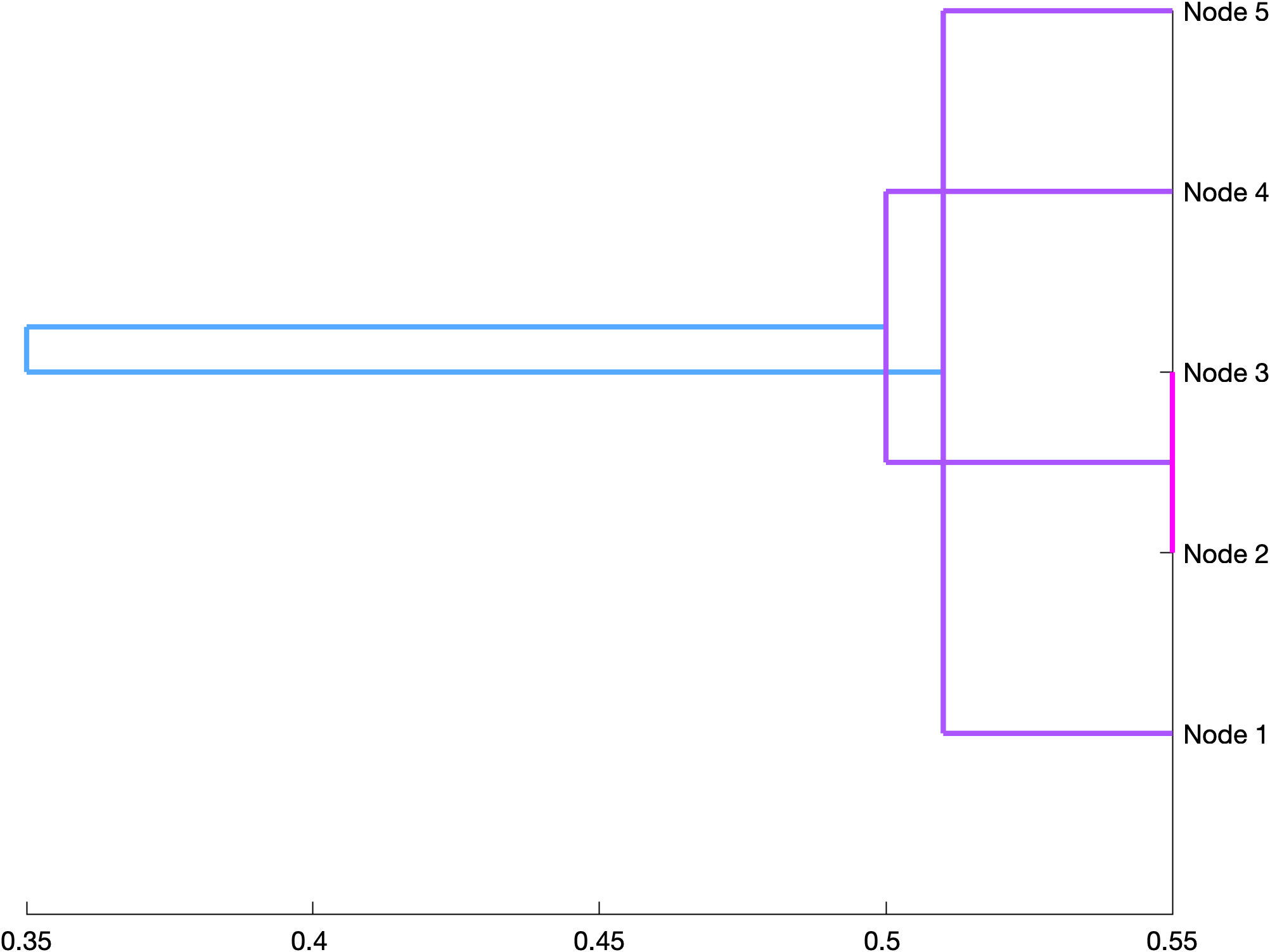}
      \caption{}
     \label{fig:ddbirth}
 \end{subfigure}
 \caption{The dendrograms for the cycles and connected components for the network in Figure \ref{fig:full}. (a) The dendrogram for cycles. The x-axis represents the filtration values (death values). (b) The dendrogram for connected components. The x-axis represents the filtration values (birth values).}
\end{figure}

 Consider the fully-connected five-node network in Figure \ref{fig:full}. We can decompose it into the maximum spanning tree (Figure \ref{fig:birth}) and the non-maximum spanning tree (Figure \ref{fig:death}). To build the dendrogram, we consider the reverse filtration process. We sequentially add the edges (smallest weighted edges first) from the maximum spanning tree to the non-maximum spanning tree. Each of this step will create a cycle. 
The binary merge is then formed by gluing the cycles sequentially. Figure \ref{fig:dddeath} shows the dendrogram corresponding to the gluing of the $1$-cycles. Figure \ref{fig:ddbirth} shows the dendrogram for the connected components.

\subsection{Isomorphic Properties of the Dendrogram}
The homology groups $\mathcal{H}_0$ and $\mathcal{H}_1$ are vector spaces. Every fixed basis vector space can be associated with a merge tree \cite{pegoraro2021metric,gueunet2017task,morozov2013distributed}.Two trees are isomorphic if they differ by either node or edge label, but are topologically equivalent, and where the topology is influenced by the functional space \cite{ta2010nonlocal}.

\begin{figure}[ht!]
	\centering
	\includegraphics[width =\textwidth]{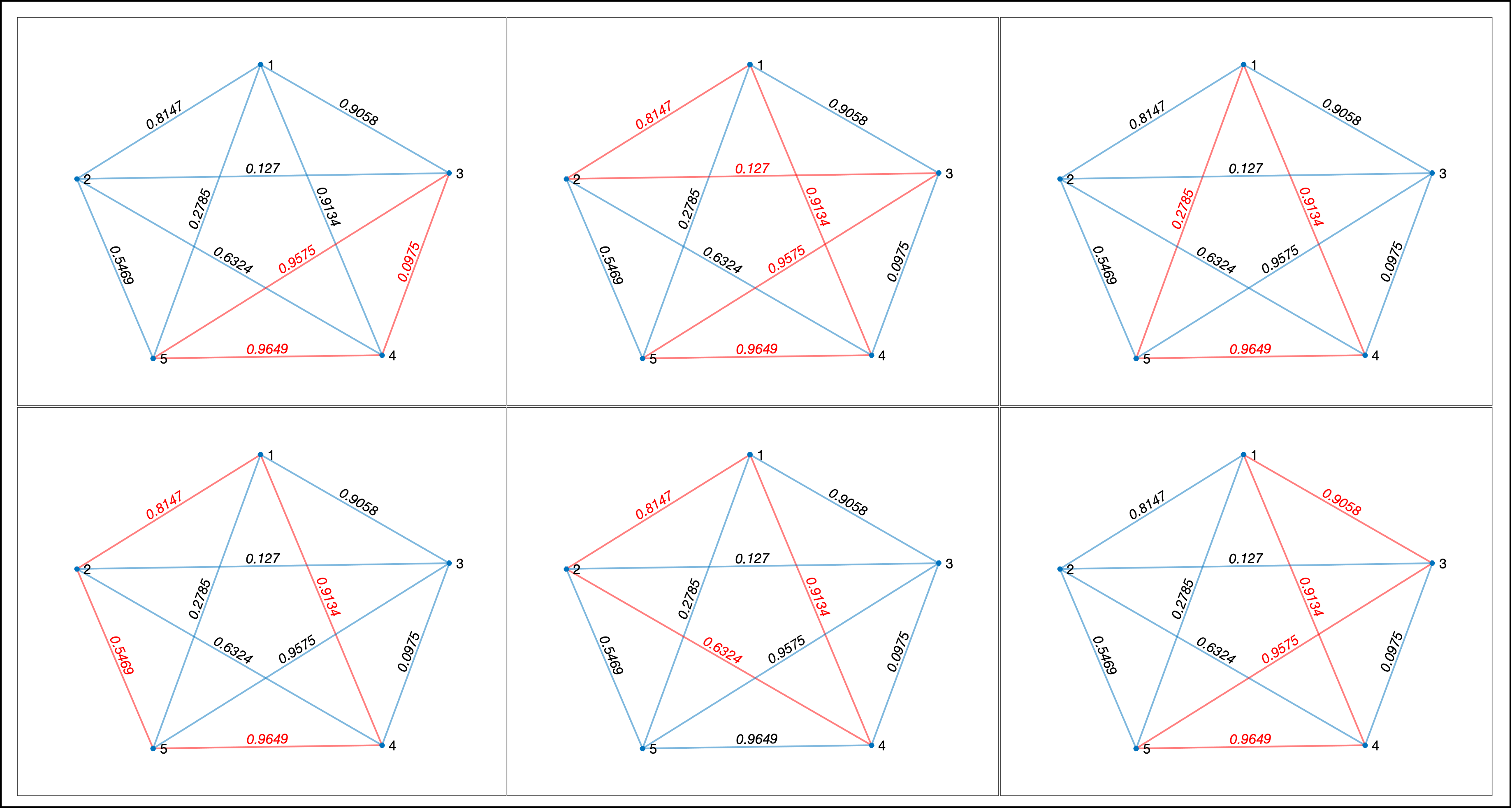}
	\caption{A fully connected five-node network with six $1$-cycles. The $1$-cycle space is identified by the red-colored edges.}
	\label{fig:cycle_space_counter_example1}
\end{figure}

\begin{figure}[ht!]
	\centering
	\includegraphics[width =\textwidth]{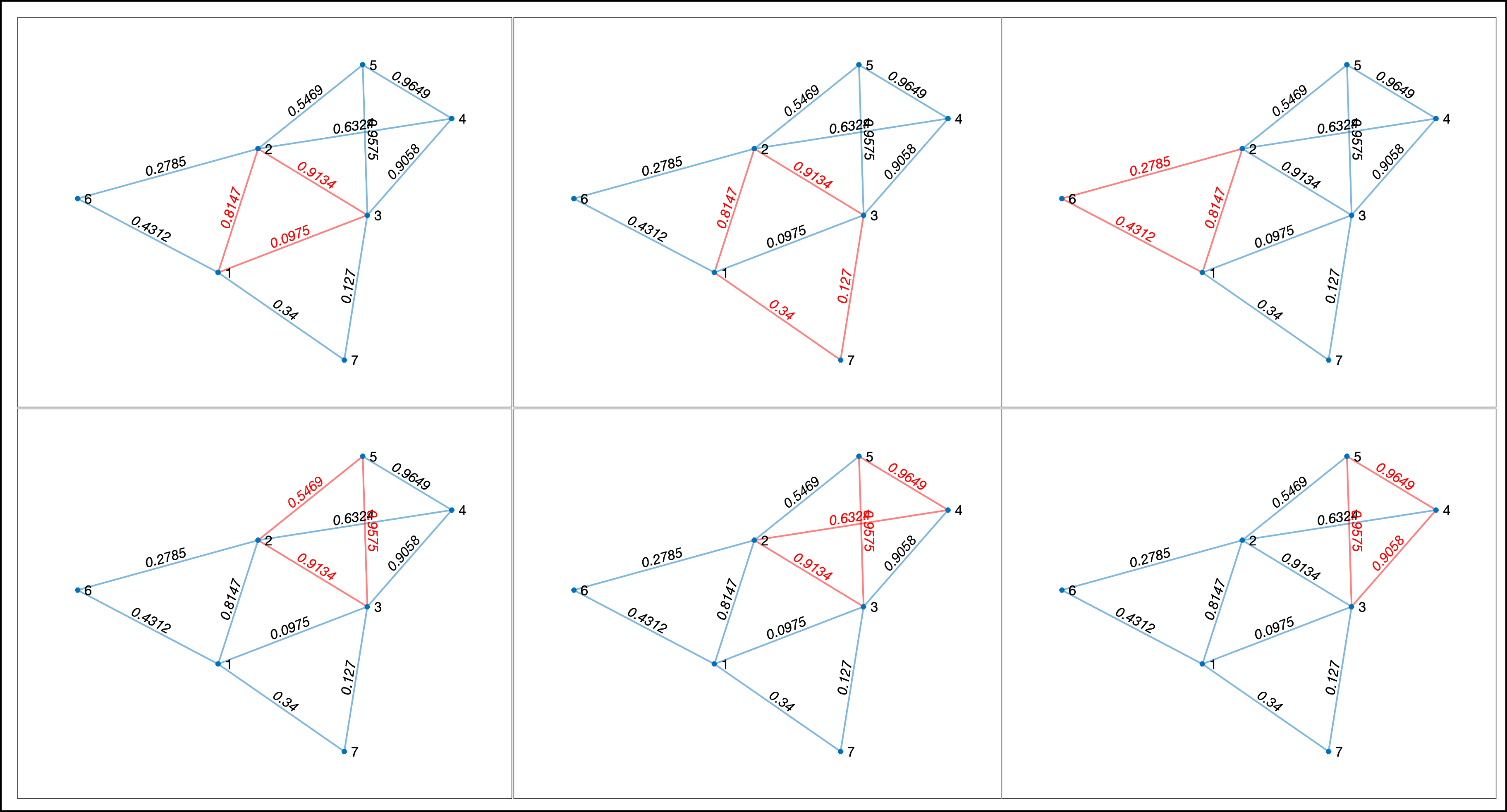}
	\caption{A seven-node network with six $1$-cycles. The $1$-cycle space is identified by the red-colored edges.}
	\label{fig:cycle_space_counter_example2}
\end{figure}

Given any two fixed basis vector spaces, their associated merge trees are isomorphic if and only if the two vector spaces are isomorphic.
Note that the vector space $\mathcal{H}_0$ is associated with the merge tree for the connected components and $\mathcal{H}_1$ is associated with that of $1$-cycles. Two dendrograms $\mathcal{D}_1 = (T_1, w_1)$ and $\mathcal{D}_2 = (T_2, w_2)$ are isomorphic if there exists a bijection $h:V_{T_1} \xrightarrow{} V_{T_2}$ such that $w_1 = h \circ w_2$
\cite{pegoraro2021metric,gueunet2017task,morozov2013distributed}. Here $w_k$ is a height function associated with the merge tree. From the foregoing exposition, it shows that if dendrograms are isomorphic on $\mathcal{H}_0$ or $\mathcal{H}_1$, there are not necessarily isomorphic on $\mathcal{H}_1$ or $\mathcal{H}_0$ respectively. To illustrate this concept, consider a fully connected five-node network in Figure \ref{fig:cycle_space_counter_example1}. It has six $1$-cycles and the $1$-cycle space is identified by the red-colored edges. Figure \ref{fig:cycle_space_counter_example2} also shows a seven-node network with six $1$-cycles and the $1$-cycle space is identified by the red-colored edges.

\begin{figure}[ht!]
\centering
\begin{subfigure}[b]{0.49\textwidth}
     \centering
     \includegraphics[width=\linewidth]{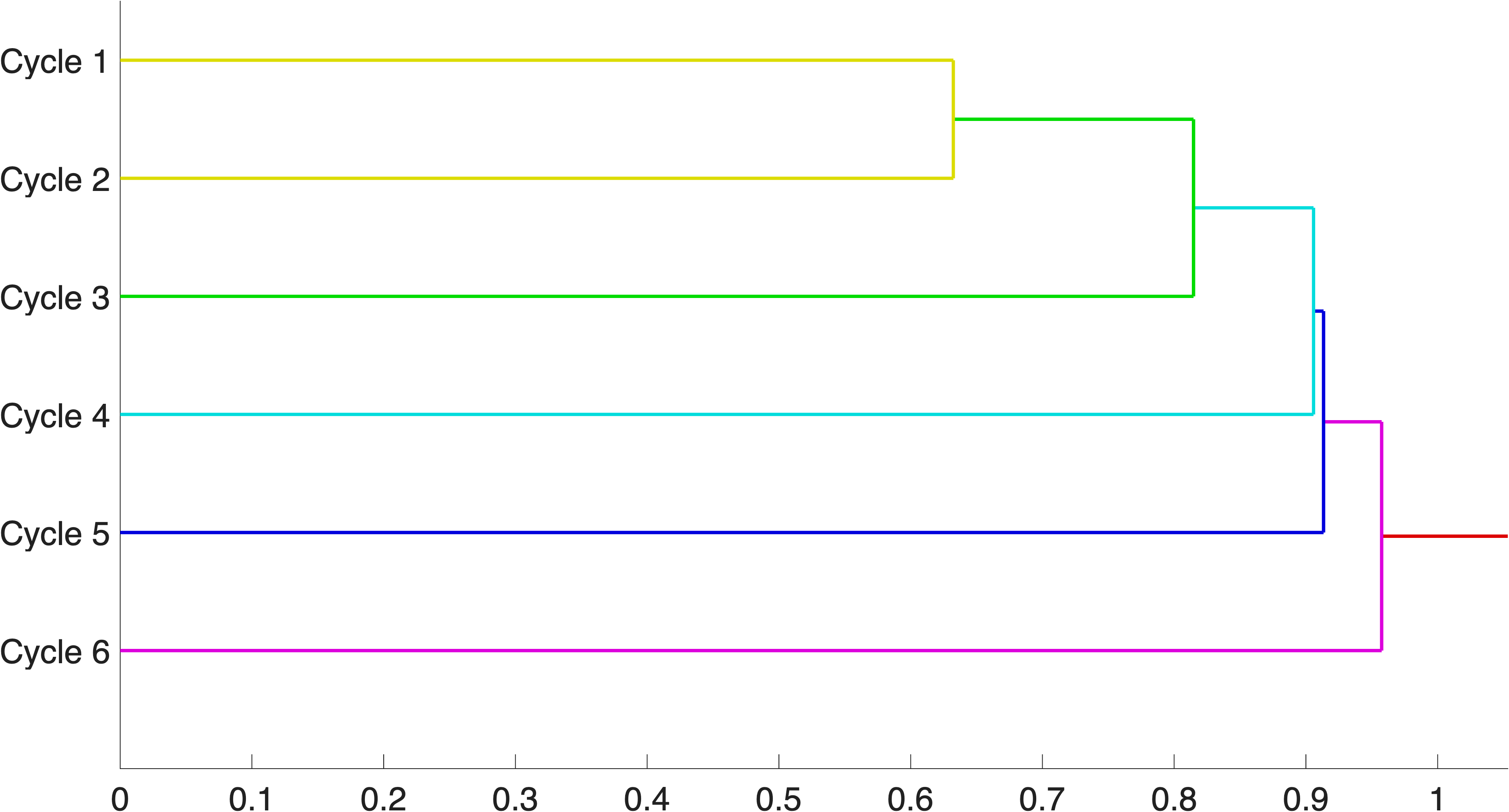}
      \caption{}
     \label{fig:dendrogram_death_5node}
 \end{subfigure}
\begin{subfigure}[b]{0.49\textwidth}
     \centering
     \includegraphics[width=\linewidth]{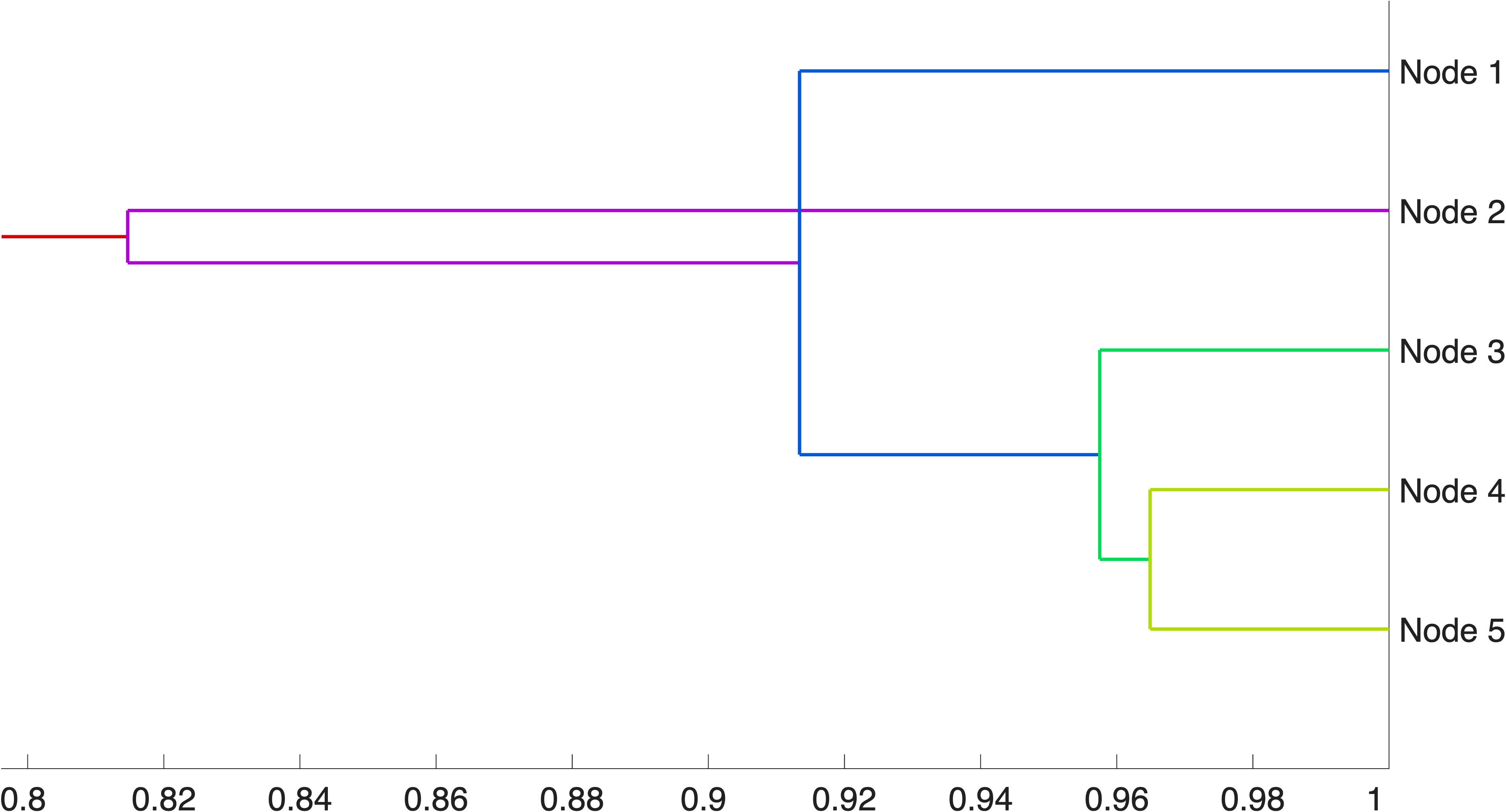}
      \caption{}
     \label{fig:dendrogram_birth_5node}
 \end{subfigure}
\vspace{0.5cm}
 \begin{subfigure}[b]{0.45\textwidth}
     \centering
     \includegraphics[width=\linewidth]{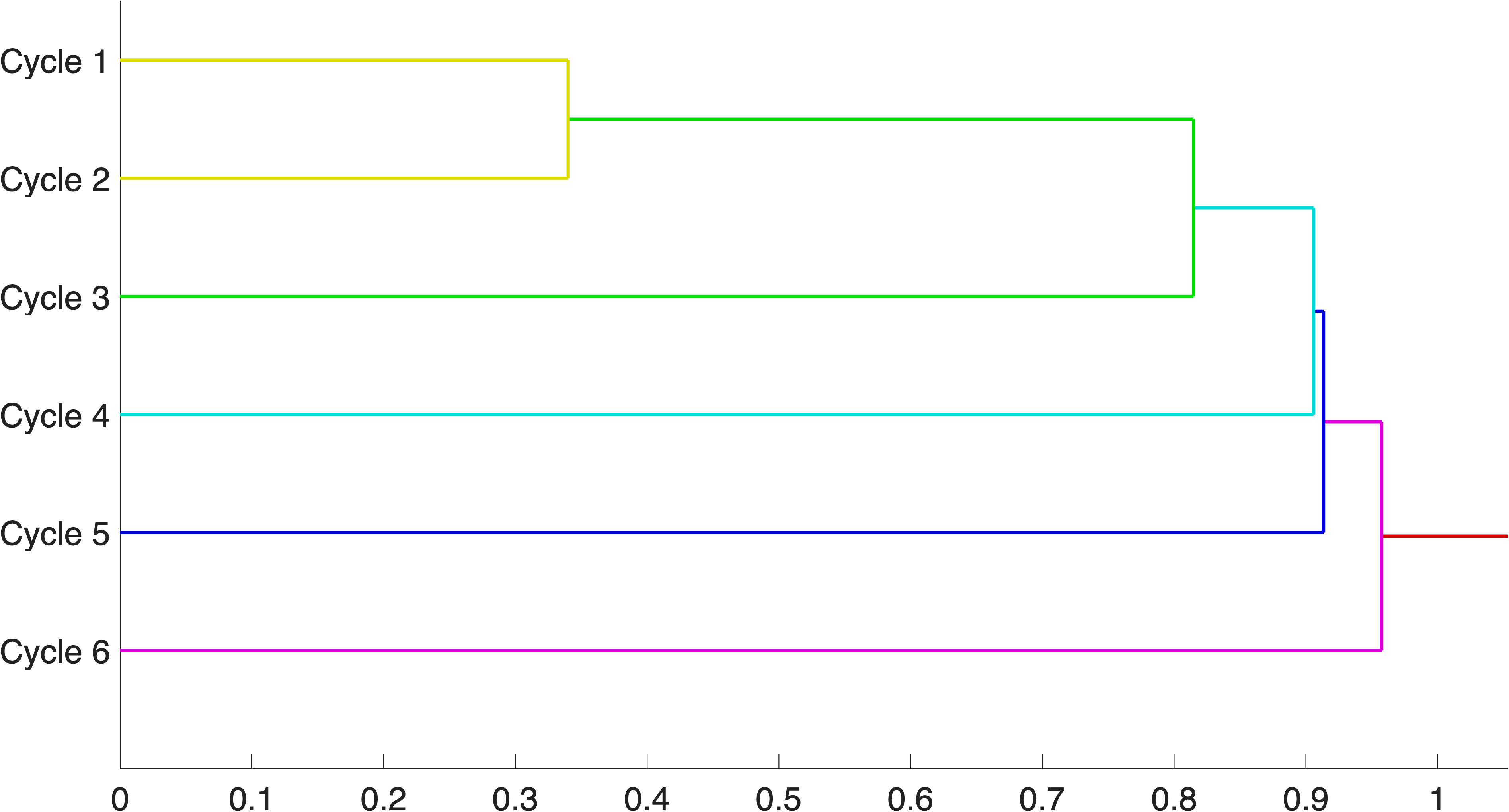}
      \caption{}
     \label{fig:dendrogram_death7node}
 \end{subfigure}
 \hfill
 \begin{subfigure}[b]{0.45\textwidth}
     \centering
     \includegraphics[width=\linewidth]{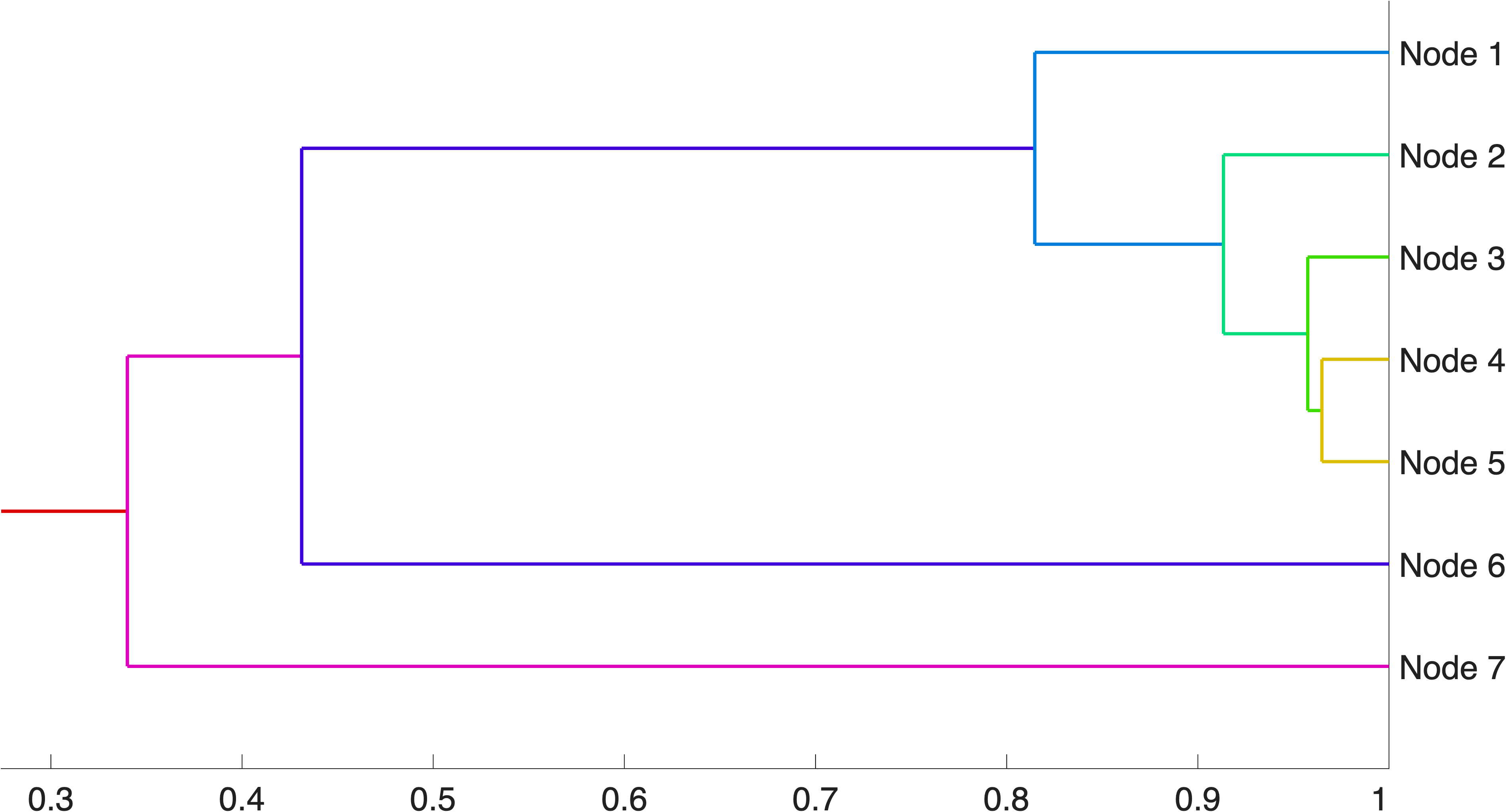}
      \caption{}
     \label{fig:dendogram_birth_7node}
 \end{subfigure}
 \caption{(a) The dendrogram corresponding to the cycle space in Figure \ref{fig:cycle_space_counter_example1}. (b) The dendrogram corresponding to the cycle space in Figure \ref{fig:cycle_space_counter_example2}. Both are identical.}
\end{figure}

 The dendrogram for persistence tree of the cycles in Figure \ref{fig:cycle_space_counter_example1} is given in Figure \ref{fig:dendrogram_death_5node}. The dendrogram for persistence tree of the cycles in Figure \ref{fig:cycle_space_counter_example2} is given in Figure \ref{fig:dendrogram_death7node}. We observe that the two dendrograms are isomorphic. However, the generating $1$-cycle space are not isomorphic, and by extension distinct 0D topology (Figure~\ref{fig:dendrogram_birth_5node} and Figure~\ref{fig:dendrogram_death_5node}). This illustrates the fact that identical dendrograms do not directly translate to isomorphism of the generating space.

However, observe that the persistence intervals $\mathbb{P}_k(K)$ are associated with the homology classes in $\mathcal{H}_k$. 
Under the assumption that the birth and death values can be perturbed to induce uniqueness, the homology group $\mathcal{H}_k$ admits a unique decomposition into $\mathbb{P}_k(K)$ \cite{edelsbrunner2008persistent,zomorodian2005computing}. That is for each element in $\mathbb{P}_k(K)$, there exist a vector in $\mathcal{H}_k$ associated with a unique k-dimensional hole. Consequently, if two complexes have identical $\mathbb{P}_k(K)$ and satisfies the uniqueness assumption, then their dendrograms will be isomorphic on $\mathcal{H}_k$ \cite{schweinhart2015statistical}.

\subsection{Distance Notion on Dendrograms}
Dendrograms are primarily characterized by three features: the cluster labels, the cluster heights and the topology of the dendrogram. The various methods developed for comparing dendrograms revolve around these three features \cite{lee2011computing,pegoraro2021metric,pont2021wasserstein}. In what follows, we discuss a distance measures for comparing dendrograms bothering on a combination of these features.

\subsubsection{Wasserstein Distance on Dendrograms}
The notion of edit distance on dendrograms has been explored in \cite{pegoraro2021metric}. The concept of edit distance on dendrograms involves finding the minimal cost required to transform one dendrogram to another with a finite sequence of edits. The Wasserstein distance on merge trees was developed following the formulation of edit distance \cite{pont2021wasserstein}. The dendrograms are an alternative but equivalent representation of merge trees. Hence one can compare dendrograms simply by comparing their merge trees using the Wasserstein distance. Here, we develop the theory necessary to compare dendrograms using the Wasserstein distance in the context of graph filtration.
We assess the topological disparity of two dendrograms using the 2-Wasserstein distance on their merge trees. Let $(\mathcal{D}_i, T_i)$ be a set of dendrogram $\mathcal{D}_i$ and the corresponding merge tree $T_i$. The persistence diagram on the merge tree is given by $\mathbb{P}_0(T_i)$, which will be denoted by $\mathbb{P}_0^i$ for brevity. For any two such sets, $(\mathcal{D}_1, T_1)$, $(\mathcal{D}_2, T_2)$, the 2-Wasserstein distance between the dendrograms is given by:
\begin{equation}
     D_{W}^2\left[\mathbb{P}_0^1, \mathbb{P}_0^2\right] = \sum_{k = 1}^p \left[b_{(k)}^1 - b_{(k)}^2\right]^2
\end{equation}
where $b_{(k)}^1$ and $b_{(k)}^2$ are the $k$-$th$ smallest birth values in $\mathbb{P}_0^1$ and  $\mathbb{P}_0^2$ respectively.  Let $\mathcal{X}$ and $\mathcal{Y}$ be two set groups of dendrograms and their associated merge trees, we test for the group difference with the following ratio statistic
\begin{equation}
    \mathcal{T}(\mathcal{X}, \mathcal{Y}) = \mathcal{L_B}/\mathcal{L_W}
    \label{eqn:dendrograms-statistic}
\end{equation}
where $\mathcal{L}_B$ and $\mathcal{L}_W$ is the average wasserstein distance for networks between the two groups and withing the two groups respectively.

\subsubsection{Comparison against Baseline Distances}

We compare the discriminative performance of the Wasserstein distance against widely used baseline methods for comparing dedndrograms. The common baseline measures for comparing dendrograms are cophenetic correlation coefficient \cite{sokal1962comparison}, the Fowlkes-Mallows statistic \cite{fowlkes1983method}, the Gromov-Hausdorff and Bottle-neck distance which are based on the single linkage distance \cite{murtagh2012algorithms,lee2011computing}. The Fowlkes-Mallows statistic do not have a standard implementation and was therefore not considered.\\ 

\begin{figure}[ht!]
\centering
\includegraphics[width=\textwidth]{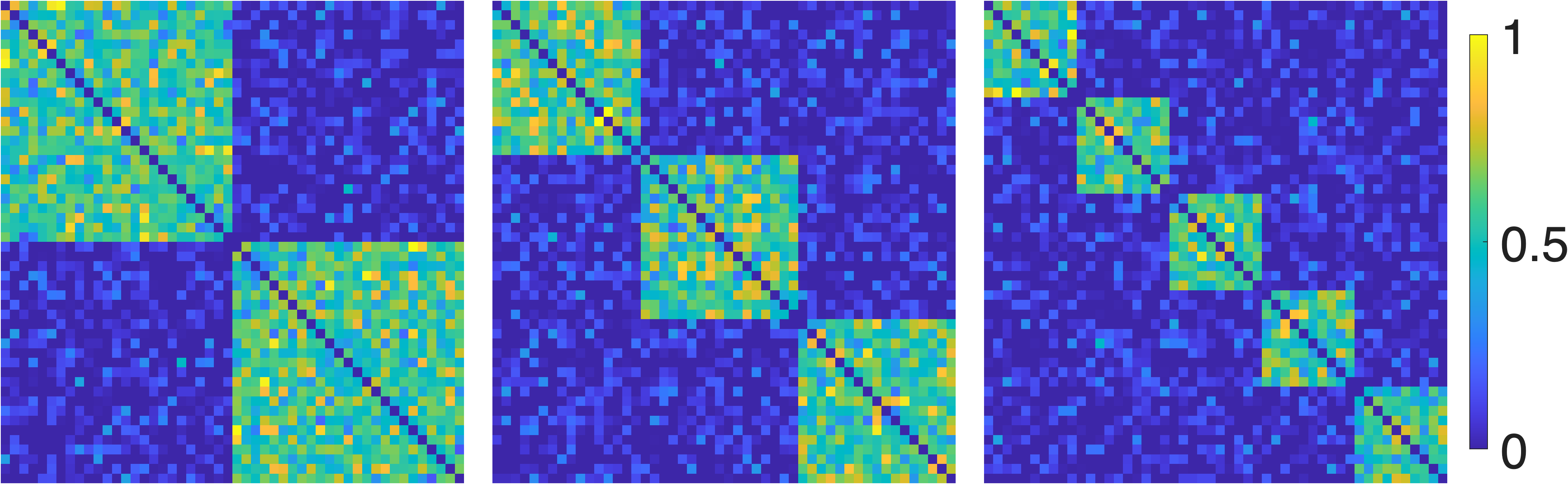}
\caption{The random modular networks. Left: The network with 2 modules. Middle: The network with 3 modules. Right: The network with 5 modules.}
\label{fig:modular_network}
\end{figure}

\noindent
\textbf{Simulation Study 1 (Connected Components):} 
We simulated three random networks with different modularity structure. We first draw a weight vector from a normal distribution with mean $1$ and standard deviation $0.25$. This weight vector was then transformed to have values in the range $0$ to $1$.  The transformed weight vector is used to form a network with a predefined modularity structure. The group $1$( Figure \ref{fig:modular_network})-left,  group $2$ (Figure \ref{fig:modular_network})-middle and group $3$ (Figure \ref{fig:modular_network})-right have $2$, $3$ and $5$ modules respectively. We fixed the intra-module connectivity probability to be $1$, and each network has 50 nodes.

\begin{figure}[ht!]
\centering
 \begin{subfigure}[b]{\textwidth}
     \centering
     \includegraphics[width=\textwidth]{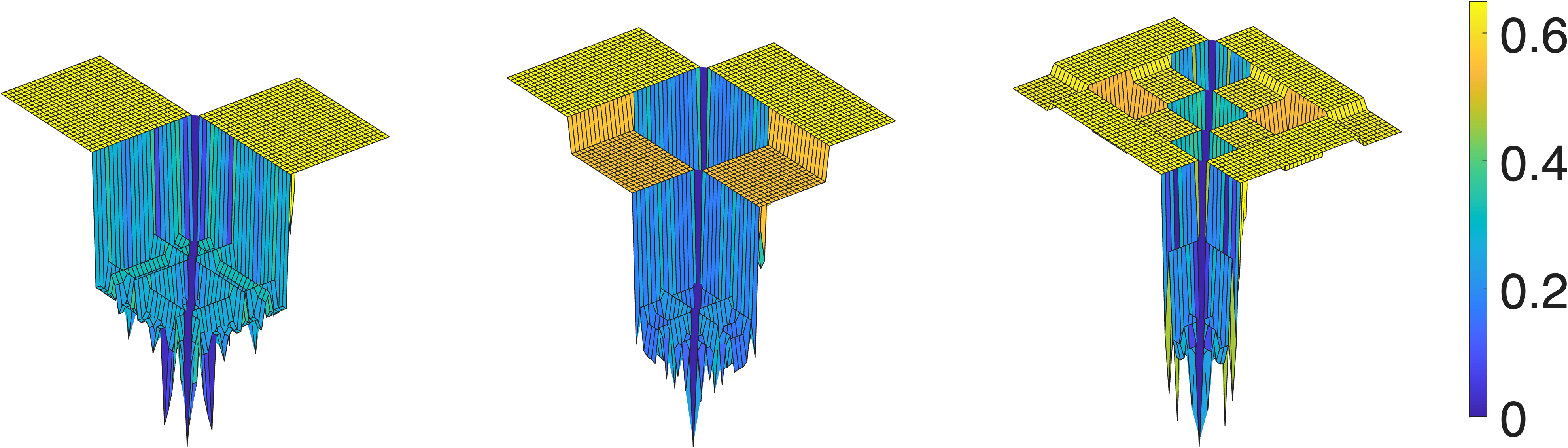}
      %\caption{Two modules}
     %\label{fig:SLM_modular_network2}
 \end{subfigure}
 %--
 %--
 \begin{subfigure}[b]{0.32\textwidth}
     \centering
     \includegraphics[width=\textwidth]{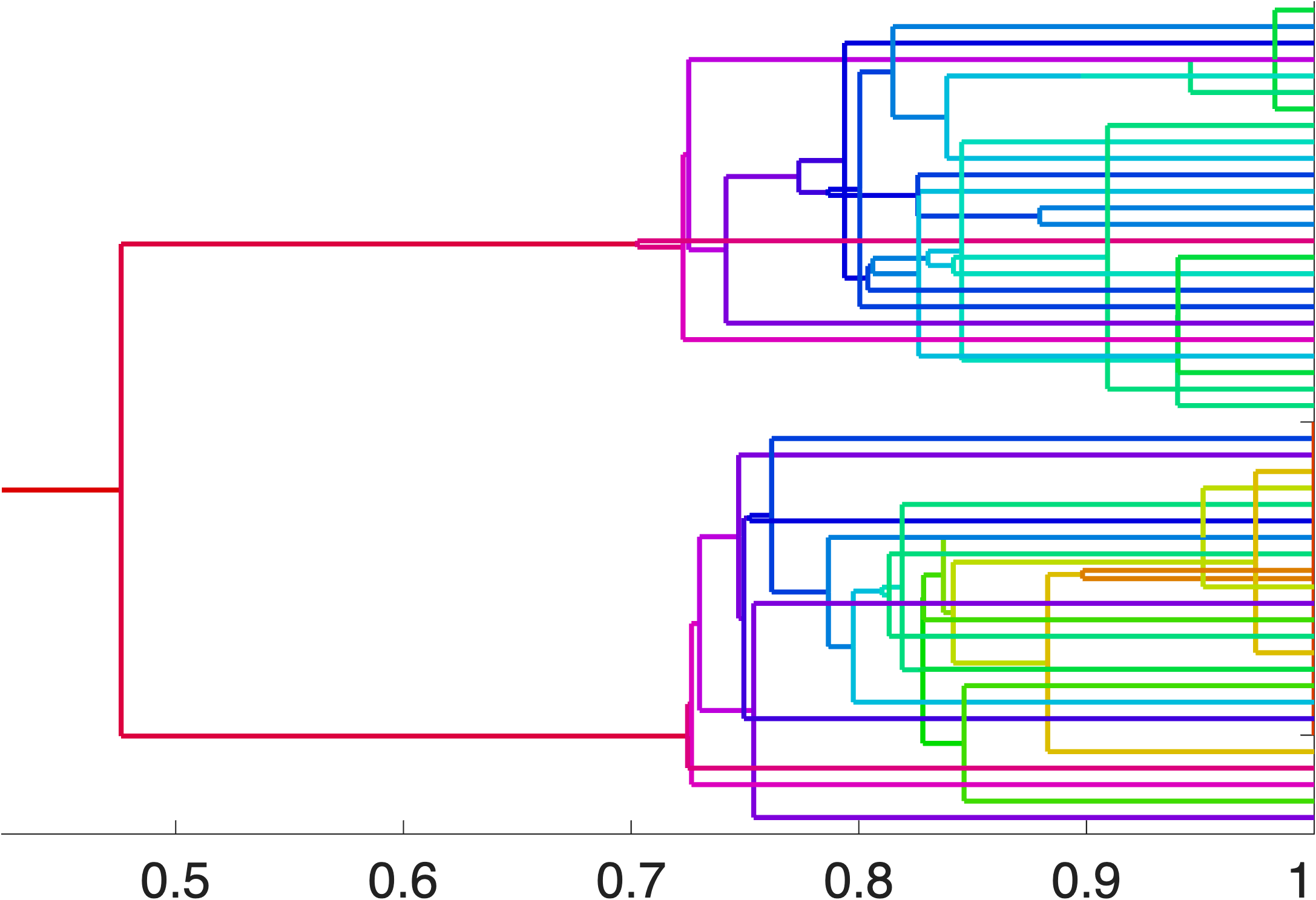}
      %\caption{Two modules}
     %\label{fig:dendrogram_modular_network2}
 \end{subfigure}
 \hfill
 \begin{subfigure}[b]{0.32\textwidth}
     \centering
     \includegraphics[width=\textwidth]{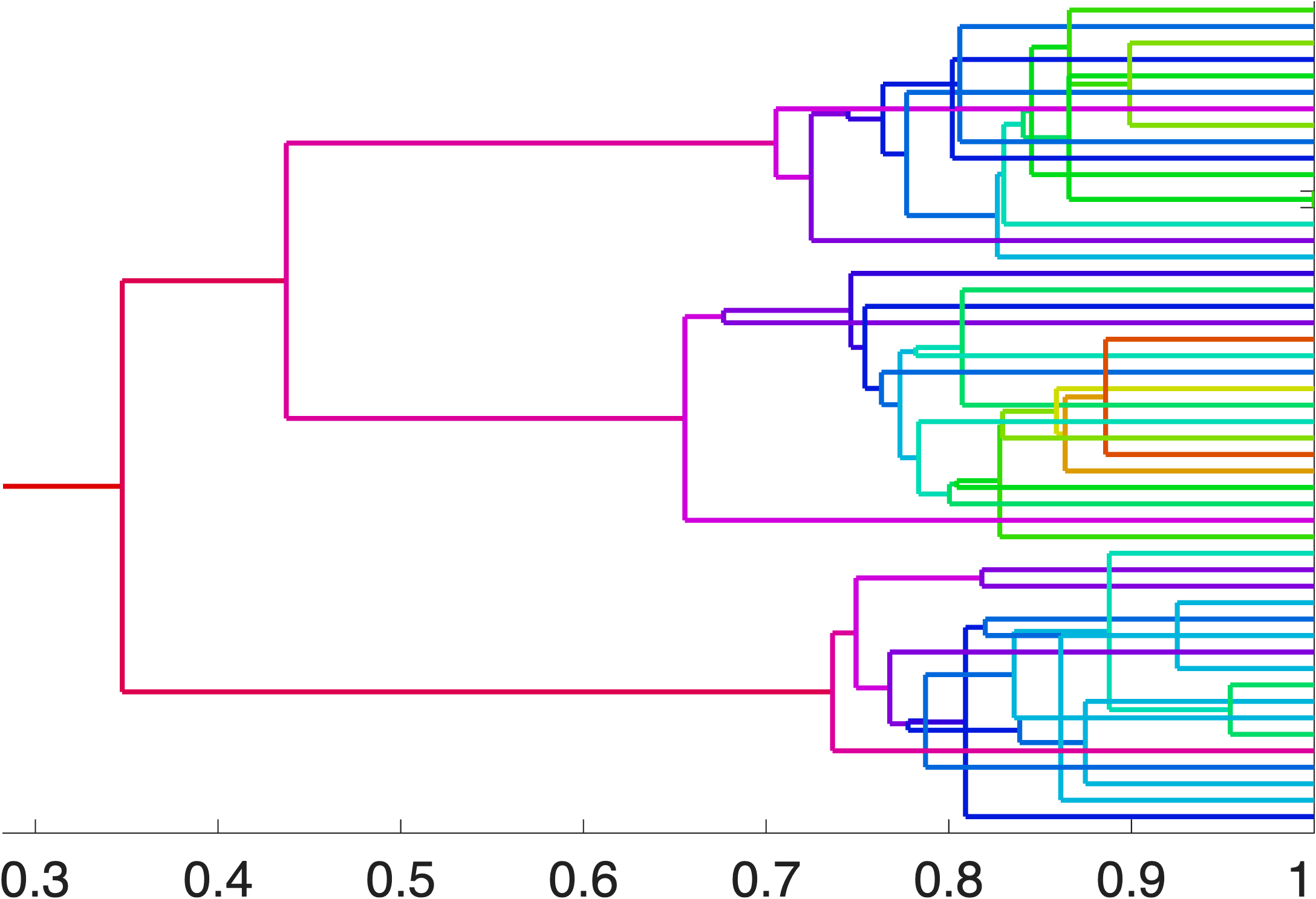}
      %\caption{Three modules}
     %\label{fig:dendrogram_modular_network3}
 \end{subfigure}
  \hfill
 \begin{subfigure}[b]{0.32\textwidth}
     \centering
     \includegraphics[width=\textwidth]{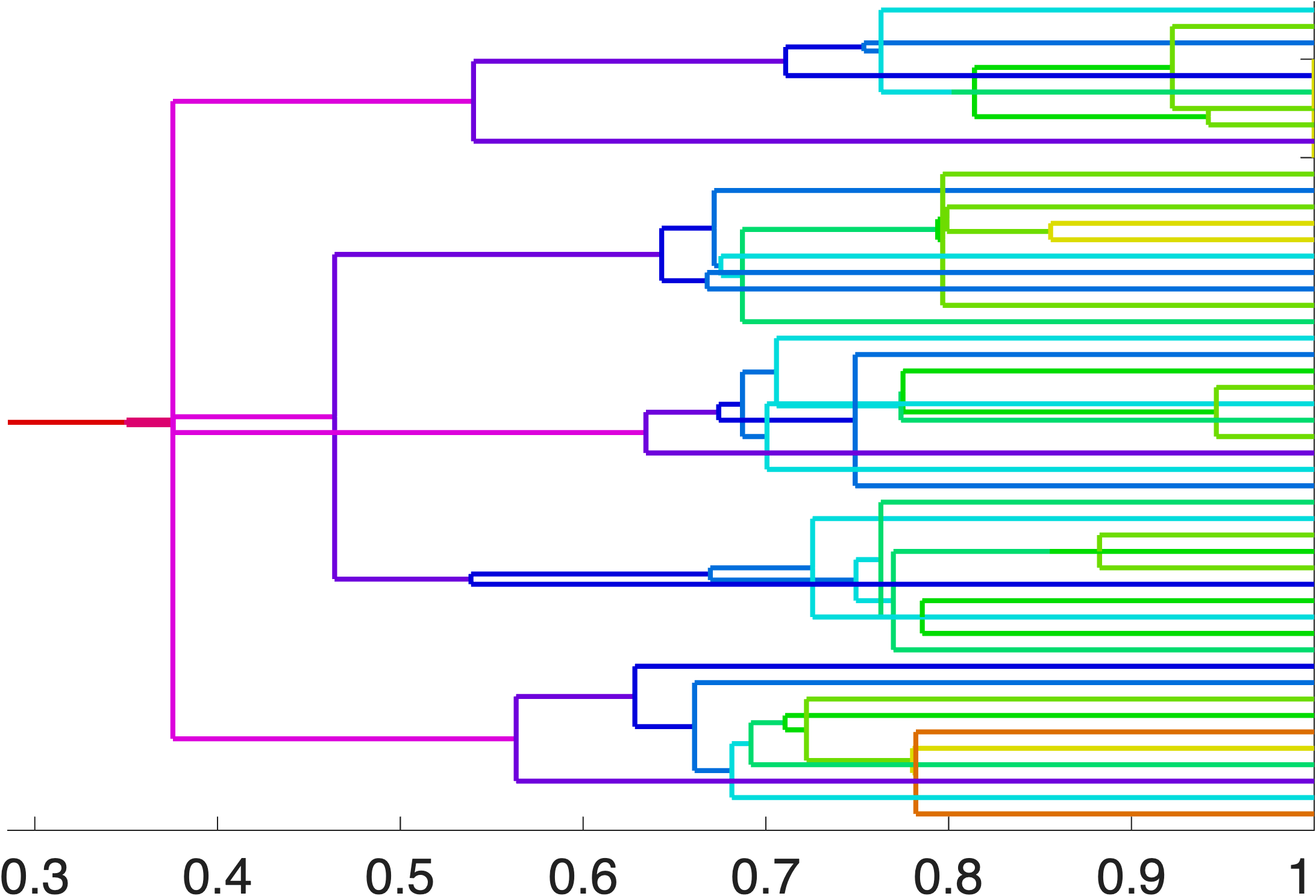}
      %\caption{Five modules}
     %\label{fig:dendrogram_modular_network5}
 \end{subfigure}
 \caption{Top: Sample SLM from the three network groups. From left:  SLM from a network with 2 modules, SLM from a network with 3 modules, SLM from a network with 5 modules. Bottom: Sample dendrogram from the three network groups. From left: dendrogram for network with 2 modules, 3 modules and 5 modules in that order.}
 \label{fig:SLM_all_networks_conncomp}
\end{figure}

\textbf{Simulation Study 1A - Clustering: } 
We carried out a cluster analysis based on the various distance measures.
We generated 10 networks in each group. The single linkage matrices (SLM) and the dendrograms of the networks in each group are then constructed. Figure \ref{fig:SLM_all_networks_conncomp} shows a sample of the SLM and dendrogram. 

We now construct the various distance matrices for clustering. We have three groups each with 10 networks. Hence we construct the $30 \times 30$ distance matrix for the Wasserstein distance using the merge trees, the Gromov-Hausdorff, Bottleneck and CP distance using the single linkage matrices. Using these distance matrices, we clustered the 30 networks into three (based on the modularity structure) clusters. We first assume the group labels of all the networks are unknown, and since we have distance matrices, we used the Ward's linkage analysis to generate the cluster labels \cite{ward1963hierarchical}. These labels were then compared to the true labels. The clustering accuracy based on the Wasserstein distance is $57\%$, the GH distance $100\%$, the Bottleneck distance $80\%$ and the CP distance $43\%$. The GH and bottleneck distance performs quite well compared to the Wasserstein and CP distance.\\

\textbf{Simulation Study 1B - Statistical Test: } We next conducted a formal test based on the three distance measures. For this, seven networks were generated from each group. The Wasserstein distance is based on the merge tree whiles the other three distances are based on the SLM. That is the observations are the merge trees and the SLMs. Each distance matrix used the ratio statistic in Equation~\ref{eqn:dendrograms-statistic}, and the statistical significance was determined using the standard permutation test with $3432$ permutations. The simulations were performed $50$ times and the  results are given in Table \ref{tab:SimulationStudy1}, where the average p-values are reported. Also, we reported the false positive rates computed as the fraction of $50$ simulations with p-values below $0.05$ and the false negative rates computed as the fraction of $50$ simulations with p-values above $0.05$. We note all the distance measures performed reasonably well with the exception of the CP distance.

\begin{table}[ht]
\renewcommand{\arraystretch}{1}
\centering
\begin{tabular}{c|c|c|c|c}
    %loops & 
    Groups & WS & GH & BN & CP \\ \hline
  \multirow{2}{*}{ 1 vs. 1} &  $0.52 \pm 0.28$ & $0.54 \pm 0.29$ & $0.55 \pm 0.32$ & $0.56 \pm 0.28$\\
  & $(0.08)$ & $(0.00)$ & $(0.00)$ & $(0.00)$\\ \hline

  \multirow{2}{*}{ 2 vs. 2} & $0.52 \pm 0.28$ & $0.43 \pm 0.26$ & $0.43 \pm 0.29$ & $0.53 \pm 0.29$\\
  & $(0.08)$ & $(0.04)$ & $(0.08)$ & $(0.00)$\\ \hline
  
  \multirow{2}{*}{ 3 vs. 3} & $0.43 \pm 0.33$ & $0.46 \pm 0.32$ & $0.37 \pm 0.30$ & $0.37 \pm 0.28$\\
  & $(0.08)$ & $(0.04)$ & $(0.16)$ & $(0.12)$\\ \hline \hline
  
 \multirow{2}{*}{ 1 vs. 2} & $0.00 \pm 0.00$ & $0.00 \pm 0.00$ & $0.00 \pm 0.01$ & $0.40 \pm 0.24$\\
  & $(0.00)$ & $(0.00)$ & $(0.00)$ & $(1.00)$\\ \hline
  
  \multirow{2}{*}{ 1 vs. 3} & $0.00 \pm 0.00$ & $0.00 \pm 0.00$ & $0.00 \pm 0.00$ & $0.35 \pm 0.32$\\
  & $(0.00)$ & $(0.00)$ & $(0.00)$ & $(0.88)$\\ \hline
  
  \multirow{2}{*}{ 2 vs. 3} &  $0.02 \pm 0.05$ & $0.03 \pm 0.06$ & $0.03 \pm 0.05$ & $0.48 \pm 0.27$\\
  &  $(0.12)$ & $(0.20)$ & $(0.20)$ & $(0.92)$\\ \hline
\end{tabular}
\caption{The performance results showing average p-values, false positive rates (first 3 rows) and false negative rates (last 3 rows). Smaller false positive and false negative rates are preferred.}
\label{tab:SimulationStudy1}
\end{table}

\begin{figure}[ht!]
 \centering
 \includegraphics[width=\textwidth]{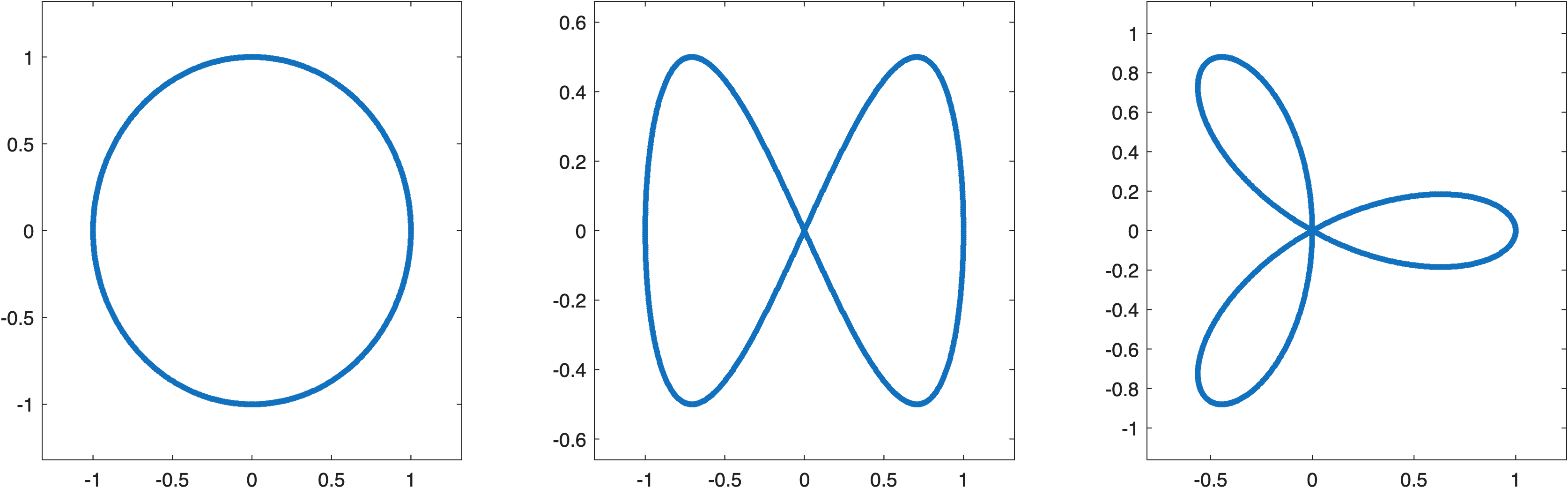}
 \caption{The three network cycle structures, with one, two and three loops. Twenty-five (25) points were sampled along the curves. These sampled points were then perturbed with  noise drawn from $N(0, 0.05^2)$.}
 \label{fig:cycle_networks}
\end{figure}

\textbf{Simulation Study 2 (1-Cycles):} This simulation investigate the clustering performance of the dendrograms. We simulated three random networks with different 1-cycle topology in a two step procedure. Group 1, 2 and 3 have one, two and three 1-cycles each respectively (Figure \ref{fig:cycle_networks}). Then 25 points were sampled along the curves. These sampled points were then perturbed with  noise drawn from $N(0, 0.05^2)$. The pairwise distance matrix of the sampled points is further normalized to be in the range of 0 to 1. In step 2, the aim is to make the dominant cycles effect more pronounced in the clustering task. To achieve this, we threshold the connectivity matrix. For any edge weight less than $0.05$, we set it to $10^{-3}\times U(0, 1)$ where $U(0, 1)$ is a random number generated uniformly on the interval $(0, 1)$.
Ten networks were generated in each group.

To perform the clustering, the distance matrices associated with the various groups of networks is computed. The Wasserstein distance and the bottleneck distance are based on the merge trees, the Gromov-Hausdorff is based on the single linkage matrices, and the CP distance is based on both the merge trees and the single linkage matrices. Using these distance matrices, we clustered the 30 networks into three clusters. We first assume the group labels of all the networks are unknown, and since we have distance matrices, we used the Ward's linkage analysis to generate the cluster labels \cite{ward1963hierarchical}. These labels were then compared to the true labels. The clustering accuracy based on the Wasserstein distance is $72\%$, the GH distance $74\%$, the Bottleneck distance $67\%$ and the CP distance $58\%$. All the distances preformed relatively well with the exception of the CP distance.\\

\section{Cycle Communities via Gradient Sampling} \label{sec:sgs}

A complimentary apptoach to the construction of cycle communities via dendrograms is the concept of stratified gradient sampling. The stratified gradient sampling approach involves developing a topological optimization process through cycle registration. The cycle registration involves building optimization over persistent homology within cycle generating strata. Variations of this concept were previously studied in \cite{dakurahregistration,leygonie2023gradient}.

\subsection{Stratified Gradient Sampling}
The concept of stratified sampling is ubiquitous in machine learning. It mostly used to create a test set in a small-sample machine learning problem. Gradeint sampling is a methodology to extend the steepest descent method of minimizing smooth functions to nonsmooth and potentially nonconvex functions. The main idea is to compute an approximate differential of nonsmooth functions by sampling gradients. When this gradient sampling is restricted to strata of a given space, it is termed the stratified gradient sampling. We introduce some of the key concepts used in developing this stratified gradient sampling methodology. 

\subsubsection{Smooth Stratifiable Functions}
In topology, stratification involves the partitioning of a topological space. Recall that our main topological objects, graphs (one-dimensional simplicial complexes) are topological spaces. The goal here is to achieve some sense of smoothness for functions defined on the various strata. More formally, consider the stratification of the topological space $K = \{ K_n \}_{n \in {N}}$. Let $f$ be a function from $K$ to the real line, and $f_n$ a restriction of $f$ to the stratum $K_n$. Then $f: K \xrightarrow[]{} \mathbb{R}$ is deemed a smooth stratifiable function if $f_n$ is twice continuously differentiable in a neighborhood of $K_n$. Other formal definitions of stratifiable functions exists, in particular, a popular one is the Whitney stratification which requires $\{K_n\}_{n\in N}$ to be Whitney \cite{bolte2007clarke}. The smoothness of $f$ restricted to each stratum guarantees that we have a unique limit of the gradient $\nabla f_n(\sigma_l \in K_n)$ given by $\nabla{K_n}f(\sigma)$ as long as $\sigma_l$ (a sequence of subsets of $K$) converges to $\sigma \in K_n$. Given these limit guarantees, we present a gradient descent algorithm for smooth stratifiable functions. 

\subsubsection{Gradient Descent on Smooth Stratifiable Functions}
The direction of steepest descent for a function $f$ with non-zero differential at a point say $\sigma$ is often denoted as $-\nabla f(\sigma)$. This can be obtained via the minimization
\begin{equation}
    \argmin_{||\mathbf{u}||_2 \le 1}\nabla f(\sigma)^\top \mathbf{u} = -\frac{\nabla f(\sigma)}{|| \nabla f(\sigma) ||_2}.
\end{equation}
Notice the slight abuse of notation here, as $f$ is not necessarily the smooth stratifiable function defined in the previous section. We will make this distinction in further expositions below. The assumption on $f$ is that it is smooth and convex. When the differentiability condition $\nabla f(\sigma) \ne 0$ is not satisfied, $-\nabla f(\sigma)$ is no longer the direction of steepest descent, and only small decreases are recorded along this direction. To provide a workaround to this, the gradient sampling solve (approximately) a min max optimization problem and obtaining the direction of descent. More formally, for a $\delta$-subdifferential of $f$ (where $f$ i snow taken to be a smooth stratifiable function) at $\sigma$ given  by $\bar{\partial}_{\delta}f(\sigma)$, the descent direction is obtained by solving the min max problem
\begin{equation}
    \min_{||\mathbf{u}||_2 \le 1} \max_{g \in \bar{\partial}_\delta f(\sigma)}g^\top \mathbf{u}.
    \label{eqn:min-max-desc}
\end{equation}
Observe that $\bar{\partial}_\delta$ is a convex set,  hence the descent direction will be a projection of the origin on this set. In particular, we have that $\langle g(\sigma, \delta), g(\sigma, \delta) -  g\rangle \le 0$ where the descent direction is obtained from (\ref{eqn:min-max-desc}) and have the explicit form
\begin{equation}
    g(\sigma, \delta) = \argmin\{ ||g||_2^2, g \in \bar{\partial}_{\delta} f(\sigma) \}.
    \label{eqn:descent-direction}
\end{equation}

For a given $\delta$-neighborhood the convex set $\bar{\partial}_{\delta} f(\sigma)$ will consists of infinitely many gradients, hence we will work with the assumption that, the gradient information in $\bar{\partial}_\delta f(x)$ are $\delta$-close to $\sigma$. The development in this section lays both the theoretical and practical foundations for optimizing persistence maps which falls in the class of nonsmooth and nonconvex functions but can be made stratifiably smooth. The optimization over these persistence maps is required for cycle community construction.

\subsection{Topological Centroid Registration}
\begin{figure}[ht]
 \centering
 \includegraphics[width=1\linewidth,clip=true]{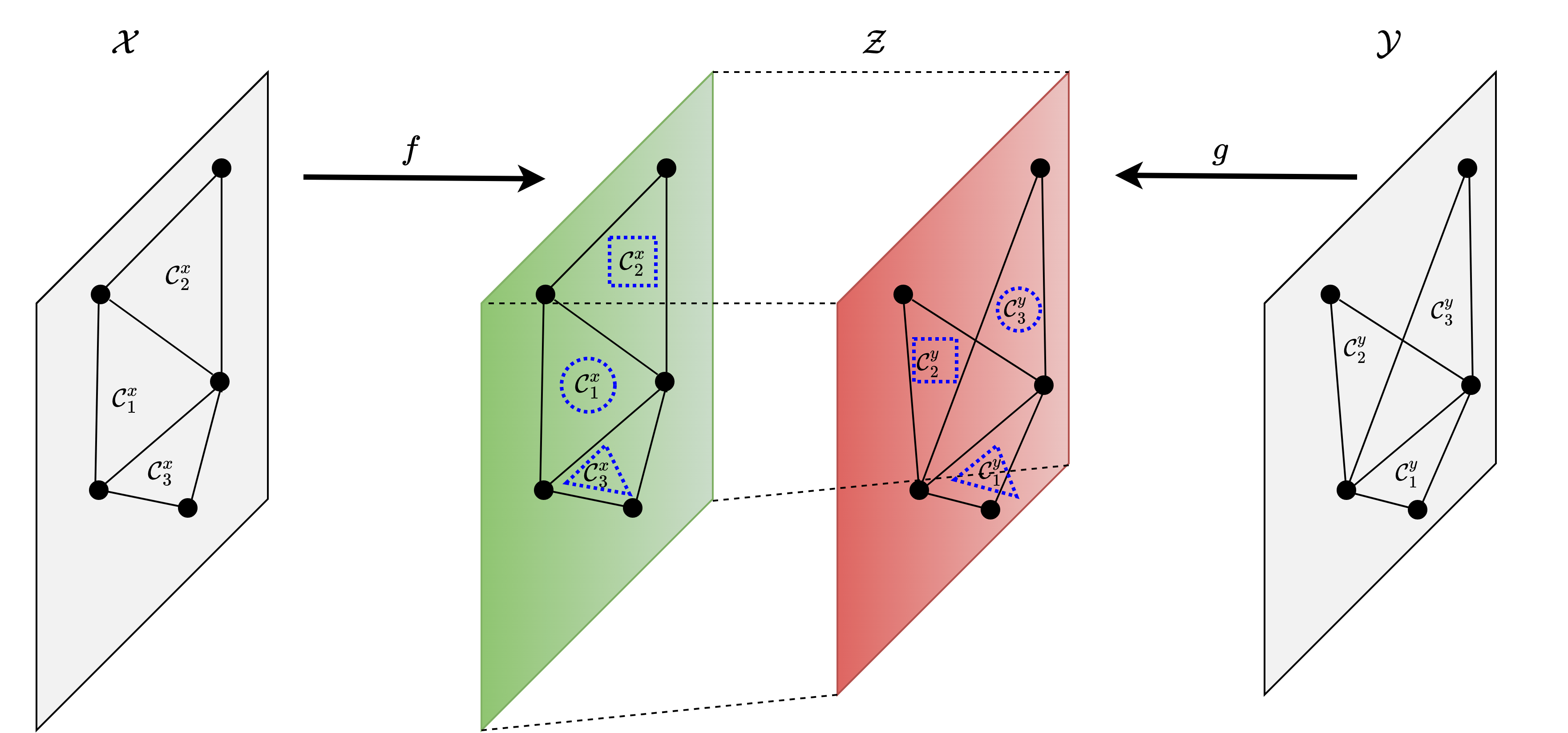}
 \caption{Illustration of the matching in the homology groups. From left is the topological space $\mathcal{X}$, the third space $\mathcal{Z}$ and the other main topological space $\mathcal{Y}$. In the space $\mathcal{Z}$, cycles enclosed in the same geometric shape are matching cycles from their respective topological spaces and they map to the same non-trivial cycle in $\mathcal{Z}$. Note that we restrict our matching to the homology basis generators.}
 \label{fig:matching}
\end{figure}

We introduce the concept of topological centroid registration. The main idea is to find a cycle centroids of one-dimensional simplicial complexes (graphs). Since each cycle is a sub-complex, the process is akin to finding the topological barycenter of a collection of subcomplexes. Figure \ref{fig:matching} provides an abstract graphical illustration of this concept.
\\

\noindent
The exposition in this section will be restricted to the one-dimensional simplicial complex $K$ and its first homology group/module $\mathcal{H}_1$. Since $\mathcal{H}_1$ is fully characterized by $D(K)$, the 1D-barcode, any reference to a barcode will be assumed to be $D(K)$. We define a persistence map $PH(.)$ that takes values from $K$ to $D(K)$ through the filter function $F$. This involves transforming $D(K)$ into a metric space by equipping it with the Wasserstein distance. Let $\phi$ be a bijective map between two barcodes $D_1$ and $D_2$ with the form $\phi: D1 \xrightarrow[]{} D_2$. The $p$-th Wasserstein distance between two diagrams has the form
\begin{equation}
    W_p(D_1, D_2) = \inf_{\phi} \left( \sum_{(b, d) \in D} ||\phi(b, d) - (b, d)||_2^p\right)^{1/p}.
\end{equation}
We observe that under the filtration approach adopted in this work, this distance admits a simplified form for $p = 2$ summarized in the proposition below.

\begin{prop}
    Let $D_1$ and $D_2$  be two sets of barcodes defined according to (\ref{eqn:barcodes}). The 2-Wasserstein distance between $D_1$ and $D_2$ admits the simplified form
    \begin{equation}
       W_2(D_1, D_2) = \inf_{\phi} \left( \sum_{(b, d) \in D} ||\phi(b, d) - (b, d)||_2^2\right)^{1/2} = 
       \left( \sum_{j = 1}^q |d^1_{(j)} - d^2_{(j)}|^2\right)^{1/2},
    \end{equation}
    where $d^1_{(j)}$ and $d^2_{(j)}$ are the $j$-th ordered values of the death times in $D_1$ and $D_2$ respectively.
\end{prop}

The proof is a direct consequence of order statistics, since by the assumption of our filtration, all the cycles are formed when the simplex is first created hence have the same birth values. This allows us to simply sort and match the death values. Following from this, we have the map $PH: K \xrightarrow[]{} D(K)$ is Lipschitz continuous, and is a result of the stability results associated with persistence diagrams \cite{edelsbrunner2008persistent}. In what follows, any discussion of the p-Wasserstein distance will be the context of $p = 2$. It remains to show that the filter functions are stratifiably smooth.\\

Consider a map from the set of barcodes $D(K)$ to the real line: $U: D(K) \xrightarrow[]{} \mathbb{R}$. The differentiability of $U$ is guaranteed by some results on the differentiability of persistence functions established in \cite{leygonie2021framework}. From this it follows that a filter function $F$ can be written as the composition of $PH$  and $U$, i.e., $F = U o PH$, and it is in fact a stratifiably smooth function from our previous discussions. This now allows us to optimize over this set of filter functions using the stratified gradient sampling method. The centroid registration can be summarized as follows.\\

Let $\mathcal{C}_K \in \mathcal{H}_1$, $k = 1, \cdots, \mathcal{Q}$ be the set of cycles, where $\mathcal{Q}$ is the cardinality of the basis cycles in $\mathcal{H}_1$. Define $\mathbb{F} = \{F_1, \cdots, F_{\mathcal{Q}}\}$ to be the set of filter functions, each corresponding to $\mathcal{C}_k$. Let $\mathcal{C}^\prime_r$, $r = 1, \cdots, \mathcal{P}$ be a set of template cycles such that $\mathcal{P} \ll \mathcal{Q}$. We can learn a set of filter functions $\mathbb{F}^\prime = \{ F_1^\prime, \cdots, F^\prime_\mathcal{P} \}$ such that
\begin{equation}
    \sum_{r = 1}^\mathcal{P} \sum_{k = 1}^\mathcal{Q} W_p\left(PH(F^\prime_r, \mathcal{C}^\prime_r), PH(F_k, \mathcal{C}_k^{(r)})\right)
    \label{eqn:objective-function}
\end{equation}
is minimized. $W_p(., .)$ is the $p$-th Wasserstein metric between the persistence intervals \cite{wasserman2018topological}, and $\mathcal{C}_k^{(r))}$ is used to denote that it can faithfully be reconstructed  from the filter function $F^\prime_r$. Since the optimization is over a filter functions which are established to stratifiably smooth \cite{dakurahregistration,leygonie2023gradient}, we are guaranteed this will converge to a stationary point. The set $\mathbb{F}^\prime = \{ F_1^\prime, \cdots, F^\prime_\mathcal{P} \}$ can be regarded as the cycle barycenters. These barycenters (centroids) can be used to form cycle communities in a manner similar to k-means clustering, and a practical implementation remains an exciting direction for future research.

\section{Discussion and Conclusion} \label{sec:dc}

This work demonstrates the novel concept of combining hierarchical clustering methods and gradient-based optimization for the novel concept of cycle community construction in topological data analysis. We have developed two complementary frameworks that together provide comprehensive tools for understanding and comparing the community structure of topological features.

The first contribution establishes a rigorous theory for dendrogram representations of homology groups. By constructing merge trees from persistence intervals using established algorithms \cite{obayashi2018volume,schweinhart2015statistical,gueunet2017task,morozov2013distributed}, we provide an interpretable hierarchical view of how cycles are related through their birth-death dynamics. The introduction of the Wasserstein distance on merge trees \cite{pont2021wasserstein,pegoraro2021metric} enables quantitative comparison of dendrograms, while our analysis of isomorphic properties clarifies when topological structures can be considered equivalent. Through simulation studies, we demonstrated the practical applications of this dendrogram-based approach, both for clustering and statistical testing tasks. These results establish dendrograms as valuable tools for cycle community detection, particularly when interpretability and hierarchical structure are priorities.

The second contribution extends the Stratified Gradient Sampling framework from learning a single filter function to simultaneously learning multiple cycle barycenter functions. The resulting cycle communities provide a natural way to partition cycles based on their topological persistence properties, analogous to k-means clustering but operating in the space of topological features. This optimization-based framework complements the dendrogram approach by offering a constructive method for learning community structures through topological registration.

The integration of these two methodologies provides a descriptive and formal tool for cycle community analysis. The dendrogram approach excels at revealing hierarchical relationships, enabling statistical comparison across groups, and providing intuitive visualizations that directly connect to hierarchical clustering theory \cite{nielsen2016hierarchical,murtagh2012algorithms,day1984efficient}. The SGS-based approach offers a principled optimization framework for learning cycle centroids and constructing communities through registration. Together, they address different aspects of the cycle community problem and can be applied either independently or in combination depending on the analytical goals.

This framework has immediate applications in neuroscience, particularly in analyzing functional brain networks where cycles represent closed loops of neural connectivity \cite{dakurah2022modelling,dakurahregistration,dakurah2025discrete,dakurah2025brain,dakurah2025spanning,dakurah2025maxtda,dakurah2025robust}, and more broadly in network analysis where identifying communities of similar cycles can reveal organizational principles and functional modules. The dendrogram representations provide neuroscientists with interpretable hierarchical structures that can be compared across subjects or experimental conditions, while the cycle barycenter approach enables registration of neural circuits to learned templates.

The theoretical foundations established here, spanning from merge-tree algorithms and dendrogram comparison to stratified optimization and topological registration, open new avenues for understanding the community structure of topological features and their role in shaping the global organization of complex networks. An important direction for future work involves investigating the relationship between dendrogram-based and SGS-based communities, particularly understanding the conditions under which these methods produce concordant or divergent results, which would provide valuable guidance for practitioners in selecting appropriate methods for specific applications. Furthermore, combining cycle community detection with other topological tools such as discrete heat kernels and Hodge Laplacians \cite{dakurah2025discrete,dakurah2022modelling} could provide deeper insights into the dynamics and functional significance of topological structures in complex systems, potentially revealing how cycle communities relate to diffusion processes, spectral properties, and harmonic structures on networks.

\bibliographystyle{plain}
\bibliography{reference}
\end{document}